\documentclass[10pt]{article}

\usepackage{amsmath, hyperref}
\usepackage{amssymb}
\usepackage{mathrsfs}
\usepackage{graphicx}
\usepackage{cite}
\usepackage[left=2.1cm,right=2.1cm,top=1.2cm,bottom=1.3cm,includeheadfoot]{geometry}
\usepackage{indentfirst}
\usepackage{color}

\begin{document}

\begin{center}
{\Large\bf A moment approach to compute quantum-gravity effects in the primordial universe}

\vskip 4mm

David Brizuela
\footnote{
E-mail address: david.brizuela@ehu.eus}
\ and
Unai Muniain\footnote{
E-mail address: unaimuni@gmail.com}

\vskip 5mm
{\sl Fisika Teorikoa eta Zientziaren Historia Saila, UPV/EHU, 644 P.K., 48080 Bilbao, Spain}

\end{center}

\begin{quotation}
\noindent
\textbf{Abstract.}\
An approach to compute quantum-gravity corrections to the scalar and tensorial power spectra
of the inflationary perturbations is presented. The analysis of the Wheeler-DeWitt equation
is performed by a decomposition of the wave function into its infinite set of moments,
which must obey certain system of (first-class) constraints.
Considering a semiclassical approximation, the system is truncated at second order in moments and
an appropriate gauge-fixing condition is introduced, which allows us to interpret the
scale factor of the universe as an internal time.
The evolution of the different fluctuations and correlations is then explicitly considered
for a de Sitter universe. An approximate analytical solution is obtained for the corrections
of the power spectra, which produces an enhancement of power for large scales. Remarkably,
the result is in agreement with previous studies in the literature that made use of
very different semiclassical approximations. Finally, the numerical implementation of the system
is also considered to verify the validity of the analytical solution.
\end{quotation}

\vskip 2mm

\noindent

\section{ Introduction}

Even if during the last decades a large amount of approaches have been attempted to construct
a consistent theory that describes the quantum behavior of the gravitational interaction
\cite{kieferquantumgravity}, the ultimate theory is yet unknown. In particular,
the lack of experimental data is one of the issues that makes progress in this
research area complicate.
Since quantum gravity effects are suppressed by the value of the Planck mass,
they are indeed very difficult to be measured. Nonetheless, these effects
could have considerable consequences on some astrophysical scenarios, such as
the primordial inflationary universe. Although the radiation from such early times is not
directly observable, quantum fluctuations present at those eras result
into anisotropies on the Cosmic Microwave Background (CMB), for which quite accurate
experimental data is already available \cite{Ade:2015xua}.

In this context, it is very usual to apply the approximation of quantum field
theory (QFT) on fixed classical backgrounds to study the evolution of such fluctuations.
In this approximation the matter and geometric perturbations are treated as quantum
entities, whereas the background universe is considered to follow a completely classical behavior.
In fact, very good results are obtained by following this method
(see, e.g., the reviews \cite{TASI1, TASI2, review3} or the textbook \cite{peter-uzan}),
which agree with the commented data.
Nevertheless it is expected that, since these quantum fluctuations were originated
at the very beginning of the universe, the quantum behavior of the background might indeed
produce some nontrivial (quantum-gravity) corrections to the standard results of
QFT on classical backgrounds.

In the present work we will analyze corrections to the power spectra of
the gauge-invariant scalar and tensorial inflationary perturbations based on the
canonical approach to quantum gravity that leads to the Wheeler-DeWitt equation.
Even if this theory might not be the fundamental theory of quantum gravity,
it is a very conservative approach and it is expected to be correct at least
at a semiclassical level.
There are several works in the literature that have already analyzed
these corrections, by using different Born-Oppenheimer-like approximations.
On the one hand, in \cite{brizuelakieferkramer1, brizuelakieferkramer2, BEKKP13, kieferkramer}
an expansion with respect to the Newton's gravitational constant $G_N$ was considered.
In particular, the term of order $\mathcal{O}(G^0_N)$ provides exactly the QFT approximation
and quantum-gravity corrections appear at next order $\mathcal{O}(G^{2}_N)$.
On the other hand, in \cite{KTV13,KTV14} a decomposition closer to the standard Born-Oppenheimer
approximation used in molecular physics was followed, by separating the wave function of
the fast degrees of freedom, which correspond to the perturbations, and of the slow ones,
given by the background universe.
In this case, the fast Hamiltonian reproduces the result
obtained within the QFT approximation, while quantum corrections are encoded in the slow sector.
Both approaches have agreed qualitatively in an enhancement of power for large scales.

The approach that will be used here is very different from these previous ones.
We will use a formalism based on quantum moments for constrained systems, as introduced in
\cite{BT09, BT10, BSST09}, which does not require a classical deparametrization.
More precisely, the wave function will be first decomposed into its infinite moments,
such as its expectation values, fluctuations and correlations.
The Wheeler-DeWitt equation will then be rewritten as an infinite set of
(first-class) constraint equations for these moments.
Since a moment of order $n$ scales as $\hbar^{n/2}$, for semiclassical peaked states
one can safely truncate the system at a given finite order and get reliable results.
In our case, in order to obtain the main quantum-gravity corrections to the inflationary power spectra, 
it will be enough to truncate the equations at second order in moments, or order $\hbar$.
After the truncation, a gauge condition will be imposed to solve the constraint system, 
that will allow us to interpret the scale factor as an internal time.
The obtained equations of motion will be solved analytically, making use of certain
approximation. In addition, the numerical implementation of the system will also be
considered, which will validate the approximate analytical solution.
The well-known results
of the QFT on classical background approximation
will be easily recovered just
by dropping the quantum moments corresponding to
the background degrees of freedom.

Finally, let us briefly mention that in the context of loop quantum cosmology,
which is also formulated by following a canonical quantization of the system,
there has been an intense activity during the last years
to obtain similar quantum-gravity corrections for the inflationary power spectra
\cite{AgMo15, SBBGLM16, BlOl16, CBMO17, ABS18, ABBMS18, MLB18, BJMM18, Agu18}.
Even if here we will not consider this theory, we point out that the method used
and developed on this paper could also be applied to that framework.

The rest of the paper is organized as follows. In Section 2 we will briefly review the classical
canonical description of an inflationary universe, which will be used to carry out the quantization in Section 3.
In this section two different quantizations will be presented: the one corresponding to the
approximation of QFT of classical background, and the full quantization of the system, that
will lead to the Wheeler-DeWitt equation. We note that, in order to perform this last quantization,
two approximations will be used: only the presence of one perturbative mode will be considered,
and the frequency of the mode will not be quantized. These assumptions
will be detailed and motivated in Section 3.2.
In Section 4 we will apply the formalism of quantum moments
for both quantization schemes and the
expression of the power spectra will be obtained.
In Section 5 we will consider the particular case of a de Sitter universe.
The usual result of QFT for the power spectrum will be derived and
corrections for that formula will be obtained by solving analytically
and numerically the equations for the moments corresponding to the Wheeler-DeWitt quantization.
Our main results and conclusions will be summarized in Section 6.

\section{The canonical formulation of the classical model}

In this section we briefly summarize the standard canonical formulation
of linear perturbations of a homogeneous and isotropic universe.

\subsection{Background homogeneous and isotropic universe}

Inflationary dynamics is very well described by considering the perturbations
of a scalar matter field $\phi$ that propagates on a homogeneous and isotropic background.
This background is described by the Friedmann-Lemaitre-Robertson-Walker metric
\begin{equation}
ds^2 = -N(t)^2 dt^2 + a(t)^2 \left(\frac{dr^2}{1-{\rm k} r^2}+r^2(d\theta^2 + \sin^2\theta d\varphi^2)\right).
\end{equation}
In this expression there are only two time-dependent variables:
the scale factor $a(t)$ and the lapse function $N(t)$.
Assuming that the matter field is minimally coupled with gravity
and its self-interaction is given by a potential $V(\phi)$,
the action for the background is given as \cite{bojowaldcanonicalgravity}
\begin{equation}
S[a,N,\phi]
= L^3 \int dt \left[ -\frac{1}{2G}\left(\frac{a \dot{a}^2}{N}-{\rm k} N\right) + \frac{a^3}{2N}\dot{\phi}^2-Na^3 V(\phi) \right], 
\end{equation}
where we have defined the reduced gravitational coupling constant as
\begin{equation}
G=\frac{4\pi G_N}{3},
\end{equation}
$G_N$ being Newton's constant, and
$L^3$ stands for the integrated spatial volume. For the spatially compact cases
$L^3$ is the total finite spatial volume of the universe; whereas for the flat
case (${\rm k}=0$), which will be considered in the rest of the paper,
it is in principle infinite. Nonetheless, in order to regularize
the above expression, it is possible to absorb this volume by rescaling the different
objects as follows,
\begin{equation}\label{rescaling}
 a\rightarrow \frac{a}{L},\quad t\rightarrow L t,\quad N\rightarrow \frac{N}{L}.
\end{equation}
After applying this rescaling, and by performing a Legendre transformation,
the Hamiltonian of this system is obtained in a straightforward way,
\begin{equation}
\mathcal{H}_0=N \left(-\frac{G\pi_a^2}{2 a} + \frac{\pi_\phi^2}{2a^3} + a^3 V(\phi)\right). \label{backgroundhamiltonian}
\end{equation}
As it is well known, the lapse function is not a dynamical variable, but
a Lagrange multiplier. The variation of the action with respect to it
provides the constraint
\begin{equation}
 \mathcal{H}_0=0,
\end{equation}
which implies that the Hamiltonian of the system must be vanishing.
From this point on, we will choose the lapse as $N=a$, which defines the
conformal time $t=\eta$.

\subsection{Cosmological perturbations}

The perturbed line element has the general form,
\begin{equation}
ds^2 = a^2(\eta) \{ -(1-2A)d\eta^2 + 2(\partial_i B) dx^i d\eta + [(1-2\psi)\delta_{ij}+2\partial_i \partial_j E + h_{ij}]dx^i dx^j \}. \label{perturb}
\end{equation}
The perturbative variables can be classified in three different
sectors: the scalar, vectorial and tensorial ones. At linear
level different sectors evolve independently and they can thus
be treated separately.

The functions $A$, $B$, $E$ and $\psi$, in combination with 
the perturbation of the field $\delta\phi$, are scalar perturbations.
Vector perturbations have not been considered when writing the above
form of the metric since, in the absence of vectorial matter,
they are purely gauge. Furthermore, this type of perturbations
are known to rapidly decay during the accelerated expansion of the universe.
Finally, $h_{ij}$ is the tensorial part that describes the gravitational
waves.

Tensorial perturbations are invariant under gauge transformations,
but the scalar perturbations are not. Nevertheless, they can be combined in a unique gauge-invariant quantity that encodes the complete physical information
of the scalar sector. This quantity is called the
Mukhanov-Sasaki master variable $v(\eta,\textbf{x})$, and is defined as \cite{MFB92}
\begin{equation}
v(\eta,\textbf{x}) := \frac{a}{a'}\left[
a'\delta\phi + 2 a'\phi'(B-E')
+ a \phi'(A + B'-E'')\right],
\end{equation}
where the prime denotes the derivative with respect to the conformal time $\eta$.
Since the equations of motion are linear, it is useful to introduce
the Fourier transform as
\begin{equation}
v(\eta,\textbf{x})= \frac{1}{(2\pi)^{3/2}} \int_{\mathbb{R}^3} d^3\textbf{k} \: v_\textbf{k}(\eta) \: e^{i\textbf{k}\cdot\textbf{x}}.\label{vFourier}
\end{equation}

In order to obtain the dynamics of the variable $v$,
the action of the system $S$ is expanded up to second order in the perturbations,
$\delta^2 S$. In terms of the Fourier components the scalar part of the action then reads
\footnote{Since the time variable $\eta$ has been rescaled above \eqref{rescaling}
to absorb the volume of the universe $L^3$,
one also needs to rescale the perturbative variable $v_{\bf k}$ (for both scalar and tensorial sectors)
and the wave number $k$ as $v_{\bf k}\rightarrow L^2 v_{\bf k}$ and
$k\rightarrow k/L$ so that $L$ does not appear in these expressions. From this point on, all the variables are rescaled
as commented. In this way, $\eta$ and $k$ are dimensionless,
whereas $a$ has dimensions of length and $v_{\bf k}^2$ has dimensions of action. For a more explicit discussion
about this rescaling transformation see, e.g., Section III of \cite{brizuelakieferkramer1}.}
\begin{equation}
\delta^2 S = \frac{1}{2} \int d\eta \int d^3\textbf{k}  \left(v'_\textbf{k} v'^{*}_\textbf{k} + v_\textbf{k} v^{*}_\textbf{k}\left[ \frac{(a\sqrt{\epsilon})''}{a\sqrt{\epsilon}}-k^2 \right]\right),
\end{equation}
where $\epsilon$ is known as the first slow-roll parameter and is defined as
\begin{equation}
\epsilon := -\frac{H'}{a H^2}, \label{epsilondefinition}
\end{equation} 
$H:=a'/a^2$ being the Hubble parameter.
From this action, it is straightforward to obtain the canonical momentum
for each $\textbf{k}$,
\begin{equation}
\pi_\textbf{k} = \frac{\partial L}{\partial v'_\textbf{k}} = v'^*_\textbf{k}.
\end{equation}
Performing then a Legendre transformation the Hamiltonian can be finally deduced:
\begin{equation}
\mathcal{H}_S = \frac{1}{2} \int d^3 \textbf{k} \; \left(\pi_\textbf{k} \pi^{*}_\textbf{k} + v_\textbf{k} v^{*}_\textbf{k}\left[k^2 - \frac{(a\sqrt{\epsilon})''}{a\sqrt{\epsilon}} \right]\right). \label{scalarhamilton}
\end{equation}
This Hamiltonian corresponds to a system of decoupled harmonic oscillators,
with a time-dependent frequency given by
\begin{equation}
\omega^2_{S} (\eta,k) := k^2 - \frac{(a\sqrt{\epsilon})''}{a\sqrt{\epsilon}}. \label{scalarfreq}
\end{equation}

Regarding tensorial perturbations, as commented above, they are described by
the tensor $h_{ij}$. Besides being symmetric,
this tensor is also traceless $h^i{}_i=0$ and divergence-free $\partial^i h_{ij}=0$.
Therefore it just contains two physical degrees of
freedom, which correspond to the two polarizations of the gravitational wave:
$h^+$ and $h^\times$. In particular, if the Cartesian $z$ coordinate is chosen to be the direction of the propagation of the transverse wave, these polarization terms are given by $h^+ = h_{xx} = -h_{yy}$ and $h^\times = h_{xy} = h_{yx}$.

The derivation of the Hamiltonian for the tensorial part is completely
analogous to the process explained above for scalars. In order to write it down,
let us define the master variable
\begin{equation}
v^s_\textbf{k} := \frac{a}{2\sqrt{3 G}}h^s_\textbf{k},
\end{equation}
where $s$ stands for the two polarizations, $+$ and $\times$. After expanding the action, transforming to Fourier components and performing the Legendre transformation, the tensorial Hamiltonian is obtained:
\begin{equation}
\mathcal{H}_T =  \int d^3 \textbf{k} \sum_s \left( \frac{1}{2}\pi^s_\textbf{k}\pi^{s*}_\textbf{k} + \frac{\omega^2_{T}(\eta, k)}{2}  v^s_\textbf{k} v^{s*}_\textbf{k} \right).\label{tensorhamilton}
\end{equation}
Note that for these tensor modes the sum for the two polarizations $s$ must be
considered. The Hamiltonian corresponds again to a system of decoupled harmonic
oscillators, but in this case the frequency is given by
\begin{equation}
\omega_{T}^2 := k^2-\frac{a''}{a}.\label{tensorfreq}
\end{equation}

Even if both $v_{\bf k}$ and $\pi_{\bf k}$ are complex,
in the Hamiltonian they always appear in quadratic combination
with their corresponding complex conjugate. In order to proceed to the quantization, for each
of the variables one in principle should define a new set of two real variables,
for instance by taking their
real and imaginary parts as done, e.g., in \cite{MVP12}.
However, since these variables never appear outside the commented quadratic combination,
and the result will thus be unaffected by this consideration,
in order to enlighten the notation we will
just use the variables $v_{\bf k}$ and $\pi_{\bf k}$ as if they were real.

\section{The quantization of the model}

In this section
two different quantizations of the classical system presented above
will be considered. The first one will be
the standard one given by quantum field theory on fixed classical
backgrounds. In this case only the perturbative variables $v_{\bf k}$ will be quantized,
whereas the background will be left classical. In the second quantization
the full model (background plus perturbations) will be quantized.

\subsection{Quantum field theory on classical backgrounds}

In this approach each mode of the tensorial and scalar perturbations can be considered
independently, and one just promotes to an operator their corresponding
Hamiltonian, either (\ref{scalarhamilton}) or (\ref{tensorhamilton}), in order to obtain the functional
Schr\"odinger equation,
\begin{equation}\label{schroedingerequation}
\hat{\cal H}_\mathbf{k} \Psi(\eta, v_\textbf{k}):=
\left(\frac{\hat{\pi}_\textbf{k}^2}{2} + \frac{\omega^2(\eta, k) \hat{v}_\textbf{k}^2}{2}\right) \Psi(\eta, v_\textbf{k}) = i\hbar\frac{\partial \Psi(\eta, v_\textbf{k})}{\partial \eta}.
\end{equation}
In this representation the momentum operator
acts as a derivative with respect to $v_\mathbf{k}$, that is
$\hat{\pi}_\textbf{k}:=-i\hbar\partial/\partial v_\mathbf{k}$, whereas
the position operator acts by multiplication.
The dependence on the background universe is entirely inside the classical
function $\omega^2(\eta, k)$, which has the form given by (\ref{scalarfreq})
or (\ref{tensorfreq}) depending whether $v_\textbf{k}$ corresponds respectively to a scalar or tensorial mode.

\subsection{Full quantization of the model: the Wheeler-DeWitt master equation}

In order to perform a quantization of the full model, one needs to consider the
total action that describes completely the background and perturbative dynamics.
This is given by
\begin{equation}
S_{total} = S + \delta^2 S.\label{totalaction}
\end{equation}
Starting from this action for a general lapse $N(t)$, the total Hamiltonian is
obtained, which is just the sum of the background and perturbative Hamiltonians:
\begin{equation}
\mathcal{H}_{total} = N(\mathcal{H}_0 + \delta^2 \mathcal{H}) = N(\mathcal{H}_0 + \mathcal{H}_S + \mathcal{H}_T).
\end{equation}
In this expression, the lapse $N$ multiplies the whole Hamiltonian
and the background constraint $\mathcal{H}_0=0$ is not satisfied anymore
but it gets quadratic corrections in the perturbative variables. The
Hamiltonian constraint now reads
\begin{equation}
\mathcal{H}_0 + \mathcal{H}_S + \mathcal{H}_T = 0.
\end{equation}
More explicitly, by choosing the lapse $N(t)=a$, which corresponds to the conformal time $\eta$,
the classical Hamiltonian constraint takes the following form,
\begin{eqnarray}
\mathcal{H}_{total} = -\frac{G}{2} \pi_a^2 + \frac{\pi_\phi^2}{2a^2} + a^4 V(\phi) +  \frac{1}{2} \int d^3 \textbf{k} \;\left(\pi_\textbf{k}^2 + \omega^2_{S} v_\textbf{k}^2 \right)
+\frac{1}{2}\int d^3 \textbf{k} \sum_{s} \left((\pi^s_\textbf{k})^2
+ \omega^2_{T} (v^s_\textbf{k})^2 \right) = 0.\label{hamiltontotal}
\end{eqnarray}

Once this Hamiltonian is at hand, the quantization procedure is performed
by promoting different variables into operators that satisfy
the canonical commutation relations,
\begin{equation}
[\hat{a},\hat{\pi}_a] = i\hbar, \qquad [\hat{\phi}, \hat{\pi}_\phi] = i\hbar, \qquad [\hat{v}_\textbf{k},\hat{\pi}_\textbf{p}] = i\hbar\delta(\textbf{k}-\textbf{p}),\label{commutators}
\end{equation}
where $v_\textbf{k}$ stands now for both scalar and tensorial components.
The action of these operators on a wave function
on the position representation $\Psi(a,\phi,\{ v_\textbf{k}\})$
is as follows,
\begin{equation*}
\hat{a}\Psi = a\Psi, \qquad \hat{\pi}_a\Psi = -i\hbar\frac{\partial \Psi}{\partial a}, \qquad 
\hat{\phi}\Psi = \phi\Psi, \qquad \hat{\pi}_\phi\Psi = -i\hbar\frac{\partial \Psi}{\partial \phi} \qquad \hat{v}_\textbf{k}\Psi = v_\textbf{k}\Psi, \qquad \hat{\pi}_\textbf{k}\Psi = -i\hbar\frac{\partial \Psi}{\partial v_\textbf{k}}. \label{canonicalquantback}
\end{equation*}

The master Wheeler-DeWitt equation is then obtained just by requesting
that the action of the Hamiltonian operator on the wave function is vanishing,
\begin{equation}\label{WdWfull}
 \hat{\cal H}_{total}\Psi=0.
\end{equation}
This equation is very difficult to deal with and, following \cite{kieferkramer},
one can further simplify it by assuming a product ansatz for the
wave function $\Psi$ and dropping cross terms that involve different modes.
In this way, one obtains an individual Wheeler-DeWitt equation for each
mode,
\begin{equation}
\hat{\mathcal{H}}\Psi_\textbf{k} := \left(-G\hat{\pi}_a^2+\frac{\hat{\pi}_\phi^2}{\hat{a}^2}+2\hat{a}^4V(\hat{\phi})+\hat{\pi}_\textbf{k}^2+\omega^2(\eta, k) \hat{v}_\textbf{k}^2\right)\Psi_\textbf{k} = 0.\label{WheelerDeWittequation}
\end{equation}
Physically, considering this last equation instead of the full one \eqref{WdWfull} corresponds to assuming just the presence of one perturbative
mode in the classical model. Therefore, for a given mode $\bf k$, we are neglecting possible effects from other modes on its dynamics, but only analyzing how
the quantum behavior of the background will affect its evolution.
This is a natural extension of the assumption performed at a classical level (and considered also in the QFT
approach) that the perturbations can be linearized, which decouples different modes, and thus their dynamics
is only influenced by the background. If now the background dynamics is modified, because of its quantum behavior,
it is reasonable to assume that each mode will be affected more importantly from the change of the background
and not from the quantum dynamics of other modes. In fact, the effects of the other modes might also come from
nonlinearities (second and higher-order terms in perturbation theory), and not only from the quantization of
the background. In this way, we focus on the essential ingredient that is neglected in the QFT approach: the
quantum behavior of the background.

The equation above is valid for both a scalar or a tensorial mode and that is why we have not written
any of the subindices $S$ or $T$ in the frequency $\omega$. In fact, from this point on,
all expressions will be valid for both scalar and tensor modes, and thus we will use the symbol
$\omega$ indistinctly.

Finally, note that in this quantization procedure the frequency $\omega$ has been left as a
classical function of time, given by its form (\ref{scalarfreq}) or (\ref{tensorfreq}).
This is a usual approximation in the literature, see e. g. \cite{brizuelakieferkramer1, KTV13}.
The main reason not to perform the quantization of the frequency in the general case is that it leads
to nonlocal operators because the classical expression of this quantity contains inverse of background momenta.
One could argue that one should follow this path and consider this nonlocality as an inherent and fundamental
feature of the model. Nevertheless, we note that the inverse of the momenta that appear inside the frequency
come from performing certain canonical transformations at a classical level in order to define and decouple the
gauge-invariant master variable from the pure-gauge and constrained variables. Nonetheless, the Hamiltonian
of the perturbations that is defined just by considering the second-order perturbation of the
usual Hamiltonian of general relativity, is linear in the momenta. Therefore, apart from the technical
difficulty that this nonlocality would add to the model, because of the commented reasons, it might not
be an actual fundamental property of the theory and might lead to spurious results.

In addition, the frequency appears in the Hamiltonian constraint multiplying the square of the perturbative
variable $v_{\bf k}$. Therefore, when expanding the Hamiltonian at second-order in moments, the fluctuations and
correlations of the frequency would also be multiplied either by $v_{\bf k}$ or $v^2_{\bf k}$. In the QFT approach, by choosing
adiabatic initial states \eqref{vpiconditions}, the variable $v_{\bf k}$ is vanishing all along evolution. For the de Sitter case,
we will show that this also happens in our framework [see equation \eqref{v0constant}]. Therefore, at second-order in moments,
the quantum effects of the frequency will, in general, be canceled out by the vanishing of the variable $v_{\bf k}$.

\section{Quantum moments and the power spectra for the perturbations} \label{sec_moments}

In this section we will analyze the two quantizations performed in the previous section
by decomposing the corresponding wave function into its infinite set of moments.

\subsection{Quantum moments for quantum field theory on classical backgrounds} \label{moments_classical_back}

In order to analyze the system given in the approximation
of QFT on curved backgrounds that is described
by the Schr\"odinger equation \eqref{schroedingerequation}, we will define
the following quantum moments,
\begin{equation}
\Delta (v_{\bf k}^n\pi_{\bf k}^m):=
\langle (\hat{v}_{\bf k}-v_{\bf k})^n
(\hat{\pi}_{\bf k}-\pi_{\bf k})^m
\rangle_{Weyl}, \label{quantummoments}
\end{equation}
where the Weyl subscript means that the expectation value is evaluated for the totally symmetric ordered
set of operators, and the expectation values $v_{\bf k}:= \langle \hat{v}_{\bf k} \rangle$
and $\pi_{\bf k}:= \langle \hat{\pi}_{\bf k} \rangle$ have been defined.
This is indeed a slight abuse of notation, as we are using the same symbols to denote
the classical variables and these expectation values. Nonetheless, we consider it is not
worth introducing a more complicate notation since the distinction will be clear from
the context and, in particular, from this point on $v_{\bf k}$ and $\pi_{\bf k}$ will
always stand for the expectation values of their corresponding operators, and not for
the classical variables.
The value $n+m$ will be referred as the order of the quantum moment.
In addition, as we will be dealing with each \textbf{k} mode separately,
we will consider that $v_\mathbf{k}$ and $\pi_\mathbf{k}$ are nonfield quantum
variables, satisfying $[\hat{v}_\textbf{k},\hat{\pi}_\textbf{k}] = i\hbar$, and
formally regularize the Dirac deltas as $\delta(0)=1$.

Following this definition, the wave function is replaced by its
infinite set of quantum moments. In fact, the knowledge of all moments
(including the expectation values $v_{\bf k}$ and $\pi_{\bf k}$) is completely equivalent to the full expression of the wave function. The advantage of this
formalism is that one works directly with observable quantities.
Furthermore, a moment of order $n$ has the dimensions of $\hbar^{\frac{n}{2}}$.
Therefore, for peaked semiclassical states ($\hbar \ll 1$), one can make
the assumption that higher-order moments are negligible, truncate the infinite
set of moments and obtain physically reliable results.

The dynamics of these quantum moments will be given by the expectation
value of the Hamiltonian $\hat{\cal H}_k$, which can
be written in terms of the above moments by performing an expansion around
the expectation values $v_{\bf k}$ and $\pi_{\bf k}$,
\begin{equation}
\langle \hat{\mathcal{H}}_k(\hat v_{\bf k},\hat\pi_{\bf k})\rangle =
{\mathcal{H}}_k( v_{\bf k},\pi_{\bf k})+\sum_{n+m\geq 2}\frac{1}{n!m!}\frac{\partial^{n+m}\mathcal{H}_k}{\partial v_{\bf k}^n
\partial \pi_{\bf k}^m}
\Delta (v_{\bf k}^n\pi_{\bf k}^m)=
\frac{1}{2}\left(\pi_\textbf{k}^2 + (\Delta \pi_\textbf{k})^2 + \omega^2 v_\textbf{k}^2 + \omega^2 (\Delta v_\textbf{k})^2\right).\label{qfthamiltonian}
\end{equation}
(Here, and in the rest of the paper,
in order to follow the standard notation, the fluctuations have been denoted as $(\Delta X)^2:=\Delta(X^2)$.)
As can be seen, in the general case the expectation value of the Hamiltonian would be given
by an infinite sum in moments. Nonetheless, when the Hamiltonian is
{\it harmonic} (defined as containing only up to quadratic combinations
of basic variables), as it is the case, only up to second-order
moments appear in the expression of its expectation value. These harmonic Hamiltonians
are of a very special kind and their properties in terms of this moment formalism are explained
in detail, for instance, in \cite{Bri14}. One of their properties
is that the equations of motion for different orders in moments decouple. In this paper
we will be interested in computing the power spectra for the perturbations
which, as will be explained below, are given by the fluctuation of
the perturbative master variable $(\Delta v_{\bf k})^2$. Therefore,
in order to obtain the exact evolution of that variable given by
this harmonic Hamiltonian, it will be enough to consider the system up to second order.

The equations of motion of all variables are obtained by computing the Poisson brackets with the above Hamiltonian.
These brackets can be defined in terms of the commutators by the standard relation $\{ \langle \hat{X} \rangle,  \langle \hat{Y} \rangle \} = -i/\hbar \langle [\hat{X},\hat{Y}] \rangle$.
In particular, the evolution for the expectation values and second-order moments are given by,
\begin{eqnarray}\label{qfteq1}
v_\textbf{k}' &=& \pi_\textbf{k},\\
\pi_\textbf{k}' &=& -\omega^2 v_\textbf{k},
\\\label{qfteq3}
\left( (\Delta v_\textbf{k})^2 \right)' &=& 2\Delta(v_\textbf{k} \pi_\textbf{k}),
\\\label{qfteq4}
\left(\Delta(v_\textbf{k} \pi_\textbf{k}) \right)' &=& (\Delta \pi_\textbf{k})^2 - \omega^2 (\Delta v_\textbf{k})^2,
\\
\left( (\Delta \pi_\textbf{k})^2 \right)' &=& -2\omega^2 \Delta(v_\textbf{k} \pi_\textbf{k}).\label{qfteq5}
\end{eqnarray}
For future convenience, we note that the equations of motion for
second-order moments can be rewritten as a unique equation for the
fluctuation $(\Delta v_\textbf{k})^2$,
\begin{equation}
\left( (\Delta v_\textbf{k})^2 \right)''' + 4\,\omega^2 \left( (\Delta v_\textbf{k})^2 \right)'+ 4\,\omega\,\omega'\, (\Delta v_\textbf{k})^2 = 0.\label{deltav2general}
\end{equation}

As can be seen, the expectation values evolve exactly as in
the classical theory. In order to solve these equations is usual to
find a region where the frequency $\omega$ tends to a constant.
In inflationary scenarios this happens at the beginning of inflation
($k\eta \rightarrow -\infty$) since the mode is well-inside the horizon
and it does not feel the effects of the curvature. In that limit
the frequency tends to the wave number $\omega\rightarrow k$ and the mode
behaves as a free mode evolving on a Minkowski background. The state of the
mode is then assumed to be stationary. This stationarity condition can
be implemented just by setting equal to zero all time derivatives
in the equations above \eqref{qfteq1}--\eqref{qfteq5}; which implies that the expectation values must be vanishing,
\begin{equation}\label{vpiconditions}
 v_\textbf{k}=\pi_\textbf{k}=0,
\end{equation}
as well as the correlation,
\begin{equation}\label{condition1}
\Delta(v_\textbf{k} \pi_\textbf{k})=0.
\end{equation}
In addition, from equation \eqref{qfteq4}, the following relation between fluctuations must hold,
\begin{equation}\label{condition2}
 (\Delta \pi_\textbf{k})^2 = k^2 (\Delta v_\textbf{k})^2,
\end{equation}
where the frequency has been replaced by its constant value $k$.
Note that, no matter what is the evolution of the frequency, the stationarity
conditions for the expectation values \eqref{vpiconditions} will be obeyed during the whole evolution.
And, as long as the frequency $\omega$ is kept constant, the conditions
\eqref{condition1}-\eqref{condition2}
on second-order moments will also be obeyed. But, the evolution of the universe
makes the frequency time-dependent and thus, when the mode begins to feel
the effects of the curvature, these conditions will cease to be valid
and the values of the fluctuations and the correlation will not be constant
anymore.

Finally the above initial conditions \eqref{vpiconditions}--\eqref{condition2}
are just 4 relations for our 5-variable system \eqref{qfteq1}--\eqref{qfteq5}. Therefore, one freedom is left to be fixed, which is related to the energy of
the initial state at the beginning of inflation. It is usual, though not necessary
and in fact several works can be found in the literature analyzing an
excited initial state, see for instance \cite{Arm07, Gan11, ALP13},
to impose that the state is on its fundamental (less-energetic) level.
This choice is usually known as the Bunch-Davies vacuum \cite{BD78}.
Replacing conditions \eqref{vpiconditions}--\eqref{condition2} in the Hamiltonian \eqref{qfthamiltonian}
and taking into account that the Heisenberg uncertainty relation,
\begin{equation}
 (\Delta v_\textbf{k})^2(\Delta \pi_\textbf{k})^2 - (\Delta(v_\textbf{k}\pi_\textbf{k}))^2 \geq \frac{\hbar^2}{4},\label{uncertainty}
\end{equation}
must be obeyed, it is easy to obtain the initial value for the fluctuations
of the basic variables,
\begin{equation}\label{bdvacuum}
(\Delta \pi_\textbf{k})^2 = \frac{\hbar k}{2},\qquad
(\Delta v_\textbf{k})^2 =\frac{\hbar}{2 k}.
\end{equation}
This obviously corresponds to the fundamental state of the harmonic oscillator and its energy is $\hbar k/2$.
As a side remark, let us mention that in this framework this is the only place where the Planck constant $\hbar$
enters the computations (it does not appear in the Hamiltonian \eqref{qfthamiltonian} or in the equations of motion \eqref{qfteq1}--\eqref{qfteq5}). Therefore,
the appearance of the Planck constant in the result will be just due to this choice of initial state.

Finally, let us note that the dynamics given by the present Hamiltonian
\eqref{qfthamiltonian} conserves the combination that appears on the left-hand side of the uncertainty relation
\eqref{uncertainty}, which is another generic property of the harmonic Hamiltonians \cite{Bri14}.
This happens because the evolution of the uncertainty relation
is given by third-order derivatives of the Hamiltonian and, thus, if those are vanishing this combination is conserved. Therefore, any state saturating initially
the uncertainty relation will saturate it during the whole evolution. Even so, in
the general case, each moment independently will not be constant and thus the state
will be deformed through evolution.

\subsection{Quantum moments for the Wheeler-DeWitt equation}

In the full quantum model, the background degrees of freedom, described
by the couples $(a,\pi_a)$ and $(\phi,\pi_\phi)$, are also promoted
to operators. Therefore, in order to describe the complete system, their
corresponding moments have also to be taken into account. The notation will
be similar to the one used in the previous section. We will only deal with
up to second-order quantities. Therefore the fluctuation of a given
operator $\hat X$ will be denoted as $(\Delta X)^2$, whereas the correlation
between two operators $\hat X$ and $\hat Y$, will be denoted as $\Delta (X Y)$,
where the operators will be symmetrically ordered and thus $\Delta (X Y)=\Delta (YX)$.

As opposed to the case of QFT on classical backgrounds
of the previous section, in the full quantum picture there
is not an evolution equation of the form \eqref{schroedingerequation}, rather
the constraint equation (\ref{WheelerDeWittequation}) governs
the dynamics of the system. Furthermore, in this case the Hamiltonian
is not harmonic and different orders in moments couple. In particular,
higher-order moments enter into the equations of motion for the
expectation values, and thus these do not follow the classical trajectories
anymore. As a first approximation, we will assume a peaked semiclassical state
and introduce a truncation of the system at second order in moments.

In order to rewrite the Wheeler-DeWitt equation \eqref{WheelerDeWittequation}
in terms of the moments we will follow \cite{BT09} and request that,
since the action of the Hamiltonian on any physical wave function
must be vanishing,
the expectation value of the Hamiltonian constraint multiplied
from the left-hand side by any product of basic operators should be
vanishing as well. More precisely, the following set of infinite
relations should be obeyed
\begin{equation}
\langle (\hat{\pi}_a-\pi_a)^{n_1}(\hat{a}-a)^{n_2}(\hat{\pi}_\phi-\pi_\phi)^{n_3} (\hat{\phi}-\phi)^{n_4} (\hat{\pi}_\textbf{k}-\pi_\textbf{k})^{n_5} (\hat{v}_\textbf{k}-v_\textbf{k})^{n_6}  \hat{\mathcal{H}} \rangle = 0.\label{towerconstraints}
\end{equation}
This is in fact a tower of constraints that should be solved order
by order. As commented, we will truncate the system at second order.
In this way, only 7 constraints will have to be satisfied, which are the following:
{\allowdisplaybreaks
\begin{eqnarray}
C &:=& \langle \hat{\mathcal{H}} \rangle =  -G\left(\pi_a^2+ (\Delta \pi_a)^2\right) + \frac{1}{a^4}\left(a^2 \pi_\phi^2 + a^2 (\Delta \pi_\phi)^2-4 a \pi_\phi \Delta(a \pi_\phi) + 3\pi_\phi^2 (\Delta a)^2\right)
\notag \\ 
&+&2a^2 V(\phi)\,\left(a^2+6 (\Delta a)^2\right)
+ 8a^3 V'(\phi) \Delta(a \phi)+a^4 V''(\phi) (\Delta \phi)^2
+ \pi_\textbf{k}^2 + (\Delta \pi_\textbf{k})^2 
\notag \\
&+& \omega^2(v_\textbf{k}^2 + (\Delta v_\textbf{k})^2) = 0,\label{quantumconstraint1}\\
C_{\pi_a} &:= &\frac{1}{2} \langle(\hat{\pi}_a-\pi_a)\hat{\mathcal{H}}\rangle  = -G\pi_a (\Delta \pi_a)^2 + \frac{\pi_\phi}{a^2} \Delta(\pi_a \pi_\phi)+
\frac{1}{2 a^3}\left( 4a^6 V(\phi)-\pi_\phi^2\right)
\left(2 \Delta(\pi_a a)-i\hbar\right)\notag \\
&+& a^4 V'(\phi)\,\Delta(\pi_a \phi) + \pi_\textbf{k}\, \Delta(\pi_a \pi_\textbf{k}) + \omega^2 v_\textbf{k}\, \Delta(\pi_a v_\textbf{k})=0, \label{quantumconstraint2}\\ 
C_{a} &:= & \frac{1}{2} \langle(\hat{a}-a)\hat{\mathcal{H}} \rangle = -\frac{G}{2}\pi_a (2 \Delta(\pi_a a)+i \hbar) + \frac{\pi_\phi}{a^2} \Delta(a \pi_\phi)+ \frac{1}{a^3}\left( 4a^6 V(\phi)-\pi_\phi^2\right) (\Delta a)^2\notag \\\label{quantumconstraint3}
&+& a^4 V'(\phi)\,\Delta(a \phi ) + \pi_\textbf{k} \,\Delta(a \pi_\textbf{k}) + \omega^2 v_\textbf{k}\, \Delta(a v_\textbf{k})=0, \\
C_{\pi_\phi} &:= & \frac{1}{2} \langle(\hat{\pi}_\phi-\pi_\phi)\hat{\mathcal{H}} \rangle = -G\pi_a \Delta(\pi_a \pi_\phi) + \frac{\pi_\phi}{a^2} (\Delta \pi_\phi)^2+\frac{1}{a^3}\left( 4a^6 V(\phi)-\pi_\phi^2\right)\Delta(a \pi_\phi)\notag \\
&+& \frac{a^4}{2} V'(\phi)\,(2\Delta(\pi_\phi \phi)-i\hbar) + \pi_\textbf{k} \,\Delta(\pi_\phi \pi_\textbf{k}) + \omega^2 v_\textbf{k}\, \Delta(\pi_\phi v_\textbf{k})=0, \\
C_{\phi} &:= & \frac{1}{2} \langle(\hat{\phi}-\phi)\hat{\mathcal{H}}  \rangle = -G\pi_a \Delta(\pi_a \phi) + \frac{\pi_\phi}{2 a^2} (2 \Delta(\pi_\phi \phi)+i\hbar)+\frac{1}{a^3}\left( 4a^6 V(\phi)-\pi_\phi^2\right)\Delta(a \phi)\notag \\
&+& a^4 V'(\phi)\,(\Delta \phi)^2 + \pi_\textbf{k} \,\Delta(\phi \pi_\textbf{k}) + \omega^2 v_\textbf{k}\, \Delta(\phi v_\textbf{k})=0 \\
C_{\pi_{\rm k}} &:= & \frac{1}{2} \langle(\hat{\pi}_\textbf{k}-\pi_\textbf{k})\hat{\mathcal{H}}  \rangle = -G\pi_a \Delta(\pi_a \pi_\textbf{k}) + \frac{\pi_\phi}{a^2} \Delta(\pi_\phi \pi_\textbf{k})+\frac{1}{a^3}\left( 4a^6 V(\phi)-\pi_\phi^2\right)\Delta(a \pi_\textbf{k})\notag \\
&+& a^4 V'(\phi)\,\Delta(\phi \pi_\textbf{k}) + \pi_\textbf{k}\, (\Delta \pi_\textbf{k})^2 + \frac{\omega^2}{2} v_\textbf{k}\, (2 \Delta(\pi_\textbf{k} v_\textbf{k})-i\hbar)=0, \\
C_{v_{\rm k}} &:= & \frac{1}{2} \langle(\hat{v}_\textbf{k}-v_\textbf{k})\hat{\mathcal{H}} \rangle = -G\pi_a \Delta(\pi_a v_\textbf{k}) + \frac{\pi_\phi}{a^2} \Delta(\pi_\phi v_\textbf{k})+\frac{1}{a^3}\left( 4a^6 V(\phi)-\pi_\phi^2\right)\Delta(a v_\textbf{k})\notag \\
& + & a^4 V'(\phi)\,\Delta(\phi v_\textbf{k}) + \frac{\pi_\textbf{k}}{2}(2 \Delta(\pi_\textbf{k} v_\textbf{k})+i\hbar) + \omega^2 v_\textbf{k}\, (\Delta v_\textbf{k})^2=0. \label{quantumconstraint7}
\end{eqnarray}
}
Note that, for convenience, the Hamiltonian operator has been assumed to be Weyl ordered. In order to impose such ordering, for both the inverse of the scale-factor
operator $\hat a$ and the generic potential of the field $V(\hat\phi)$ that appear inside \eqref{WheelerDeWittequation},
a Taylor expansion around their corresponding expectation values has been considered.

In the Appendix A all Poisson brackets between these different constraints are explicitly computed
up to linear order in moments or order $\hbar$. As can be seen, all of them are a linear combination
of constraints and thus vanish on shell. Therefore, all the above constraints are first class.
In summary, we have a 27-dimensional space, composed by the 6 expectation values
$(a,\pi_a,\phi,\pi_\phi,v_\mathbf{k},\pi_\mathbf{k})$, their 6 corresponding fluctuations,
and their 15 independent correlations.
The evolution of the system will be described by curves on that space
and we have the gauge freedom to choose the time that will parameterize
those curves. In principle there are no restrictions for such a choice
and any combination of the expectation values, or higher-order moments,
could do the job. Nonetheless, considering the physical system we are describing,
it is natural to choose one of the background variables as time, so that
the deparametrized system still makes sense for the purely background case
without perturbations. There are 4 such variables $(a,\pi_a,\phi,\pi_\phi)$
but, in the next section, this formalism will be applied to the de Sitter universe,
where the matter field $\phi$ is a constant and not a dynamical variable anymore.
(In such a case, an evolution in $\phi$ would describe the nonphysical evolution
between different de Sitter universes.) Hence, we will impose a gauge so that
$a$ will be our internal time variable and its canonical conjugate momentum $\pi_a$ will
play the role of a physical Hamiltonian. Let us explain in more detail
how this is actually derived from the above constraint equations.

Since we have decided to use $a$ as our time variable,
the above system of 7 equations must be understood to be solved
for all 7 objects related to its momentum $\pi_a$; namely,
$\pi_a$ itself, its fluctuation $(\Delta \pi_a)^2$,
and its 5 correlations with the rest of the variables $\Delta(X\pi_a)_{X \neq \pi_a}$.
In particular, we note that this system is linear in the commented fluctuations and correlations.
Therefore, it is straightforward to obtain the solution for the commented
second-order moments $\Delta(X\pi_a)$ from \eqref{quantumconstraint2}--\eqref{quantumconstraint7} and replace it in the expectation value of
the Hamiltonian constraint \eqref{quantumconstraint1}. In this way,
one ends up with a fourth-order polynomial equation for the momentum
$\pi_a$:
\begin{equation}\label{eqpia}
- G^2 \pi_a^4+ G A_2 \pi_a^2+i \hbar G A_1\pi_a+A_0=0,
\end{equation}
where
the coefficients $A_i$ are real functions of expectation values and moments
unrelated to $\pi_a$, that is, moments of the form $\Delta(XY)_{X\neq \pi_a, Y\neq \pi_a}$.
In particular, the only imaginary term appears as a multiplicative
factor of the linear term in $\pi_a$, and its corresponding coefficient
$A_1$ has the form,
\begin{equation}
 A_1=-\frac{\pi_\phi^2}{a^3}+4 a^3 V(\phi).
\end{equation}
Note that this coefficient is given in terms of expectation values and
does not contain any higher-order moment. In fact,
it is proportional to the ``time'' ($a$-) derivative of the classical constraint
${\cal H}_0$. 
That is, if the Hamiltonian was time ($a$-) independent, this term would not be
present and the equation to solve for $\pi_a$ would be real.
The appearance of imaginary parts for the time variable
is a generic feature of this approach, as has been analyzed in \cite{BHT11,BHT11a,HKT12,BoHa18}.
Nonetheless, as will be explained below, in our case the evolution of the universe
will make the imaginary terms to be decreasing and thus will be safe to drop them.
Furthermore, the other two coefficients of the polynomial equation are given by,
\begin{eqnarray*}
A_0&=& 2 a \Delta(a\phi) V'(\phi ) \left[\pi_\phi^2-4 a^6 V(\phi )\right]
-\frac{(\Delta a)^2}{a^6}\left[\pi_\phi^2-4 a^6 V(\phi
   )\right]^2-a^8 (\Delta \phi)^2 V'(\phi )^2
   \\
   &+&\frac{\Delta(a\pi_\phi)}{a^5}\left[2 \pi_\phi^3-8 a^6 \pi_\phi
   V(\phi )\right]-\frac{\pi_\phi^2}{a^4}(\Delta \pi_\phi)^2-2 a^4 \pi_{\bf k}
   \Delta(\phi\pi_{\bf k}) V'(\phi )-2 a^4 \omega ^2 v_\mathbf{k} \Delta(\phi v_{\bf k}) V'(\phi )
   \\
   &-&\frac{2\pi_\phi}{a^2}\left[ \pi_{\bf k} \Delta(\pi_\phi\pi_{\bf k})
   -\omega ^2 v_\mathbf{k} \Delta(\pi_\phi v_{\bf k})\right]
  - 2 a^2 \pi_\phi \Delta(\phi\pi_\phi) V'(\phi )
\\
   &+&\frac{2}{a^3}[\pi_{\bf k} \Delta(a\pi_{\bf k})
   +\omega ^2 v_{\bf k}\Delta(a v_{\bf k})] \left[\pi_\phi^2-4 a^6 V(\phi )\right]
-\pi_{\bf k}^2 (\Delta \pi_{\bf k})^2
   -2 \pi_{\bf k} \omega^2 v_{\bf k} \Delta(v_{\bf k}\pi_{\bf k})
   -\omega ^4 v^2 (\Delta v_{\bf k})^2,\\
A_2&=&
\frac{\pi_\phi^2}{a^2}+2 a^4 V(\phi)+\pi_\textbf{k}^2+\omega ^2 v_\textbf{k}^2
+\frac{3 \pi_\phi^2}{a^4} (\Delta a)^2+\frac{1}{a^2} (\Delta \pi_\phi)^2-\frac{4\pi_\phi}{a^3} \Delta (a \pi_\phi)+ 12 a^2 V(\phi )(\Delta a)^2
   \nonumber\\
   &+&a^4
   (\Delta \phi)^2 V''(\phi )
   +8 a^3 \Delta(a \phi) V'(\phi) 
   +\omega^2 (\Delta v_\textbf{k})^2+(\Delta \pi_\textbf{k})^2.
\end{eqnarray*}
Note that if $A_0,A_1\ll A_2$, the momentum would take the form $\pi_a\approx \sqrt{A_2/G}$.
That is in fact the solution one would obtain from the expectation value
of the Hamiltonian constraint \eqref{quantumconstraint1} by considering that the system
was on a ``momentum eigenstate'' and thus its fluctuation was vanishing $(\Delta \pi_a)^2=0$.
Therefore, in this sense, the coefficients $A_0$ and $A_1$ contain the effects of considering
the quantum behavior of this variable.

In fact, in our case, the term $A_1$ can be neglected. 
We will consider the de Sitter case with an initial Bunch-Davies vacuum
and, in principle, it is expected that the system will not depart too much from
its classical behavior. Therefore, the Hubble factor $H$ would be approximately
constant and the scale factor will be given by an exponential of the cosmological time $t$ as $a\approx e^{H t}$.
The rest of the variables classically go as $\pi_a\approx e^{2 H t}$, $\pi_\phi\approx 0$,
$V(\phi)\approx H^2$, $v_\textbf{k} \approx \pi_\textbf{k}\approx 0$. With this information at hand,
we obtain that the two first terms of the equation \eqref{eqpia} behave as
\begin{equation}
 \pi_a^4\approx A_2 \pi_a^2\approx e^{8 H t},
\end{equation}
whereas the linear term also increases in time, but much slower, $A_1\pi_a\approx e^{5 H t}$. Therefore, in this
case, one can neglect this last term. As a side remark, we
will keep $A_0$, even if its approximate behavior is given by $A_0\approx e^{6 H t}(\Delta a)^2$
and, once $a$ is chosen as time and its fluctuations requested to be vanishing, it will be quite small.
In summary, by dropping the linear term $A_1=0$, relation \eqref{eqpia}
is a quadratic equation for $\pi_a^2$, which can readily be solved,
\begin{equation}\label{quadraticeqpia}
\pi_a^2=\frac{1}{2G}\left(
A_2\pm \sqrt{A_2^2+4 A_0}
\right).
\end{equation}

In order to choose the correct sign inside the square root one can replace this solution,
along with the form for the correlations $\Delta(X\pi_a)$,
obtained from \eqref{quantumconstraint3}--\eqref{quantumconstraint7}, in \eqref{quantumconstraint2} and solve that relation for the fluctuation $(\Delta\pi_a)^2$.
It can be checked that, by choosing the minus sign in the expression above, and linearizing
the result in moments, the fluctuation $(\Delta\pi_a)^2$ turns out to be divergent;
whereas for the plus sign the form of this fluctuation is given by an expression linear in other second-order moments.
Therefore, in order to be consistent with the semiclassical expansion performed in this formalism,
it is necessary to choose the plus sign. In addition, the global sign that comes from taking
the square root of both sides in \eqref{quadraticeqpia} is chosen as negative in order to
get below a positive physical Hamiltonian, which will be defined as $-\pi_a$.
In this way, one ends up with the solution,
\begin{equation}
\pi_a=-\frac{1}{\sqrt{2G}}\left(
A_2+ \sqrt{A_2^2+4 A_0}
\right)^{1/2} =: -{\cal H}_a.\label{quantumhamiltonian}
\end{equation}

At this point we have already solved the constraints for the moments related to
the momentum $\pi_a$, but we have not imposed any gauge and thus there is still
some freedom left. In order to discuss and fix this gauge freedom, let us rewrite the constraints
\eqref{quantumconstraint1}--\eqref{quantumconstraint7} in the following form:
\begin{eqnarray}
C_1 &:= & \pi_a + \mathcal{H}_a = 0,\\
C_2 &:= & \Delta(\pi_a a) - f_2 = 0,\\\label{constraintc3}
C_3 &:= & (\Delta \pi_a)^2 - f_3 = 0,\\
C_4 &:= & \Delta(\pi_a \pi_\phi) - f_4 = 0,\\
C_5 &:= & \Delta(\pi_a \phi) - f_5 = 0,\\
C_6 &:= & \Delta(\pi_a \pi_\textbf{k}) - f_6 = 0,\\\label{constraintc7}
C_7 &:= & \Delta(\pi_a v_\textbf{k}) - f_7 = 0,
\end{eqnarray}
where the functions $f_i$ only depend on the expectation values
and moments not related with $\pi_a$. Note that for writing
$C_1$ a fourth-order polynomial equation has been solved and
the corresponding solution chosen as explained above. Nonetheless,
the rest of the constraints $C_2-C_7$ corresponds to linear
combinations of constraints \eqref{quantumconstraint2}--\eqref{quantumconstraint7}
and are thus completely equivalent to those. Since that system of constraints
was proven to be first-class up to linear order with respect to second-order quantum
moments, the same applies to the last system $C_1-C_7$.

Therefore, we have a system of 7 first-class constraints, and following
the standard approach to constraint systems, the generalized
Hamiltonian is given as
\begin{equation}\label{totalham}
C_{total} = \sum_{i=1}^7 \mu_i C_i = 0,
\end{equation}
with arbitrary Lagrange multipliers $\mu_i$.
Now, we need to impose a gauge and choose the Lagrange multipliers
$\mu_i$ that keeps the gauge conditions invariant.

The above constraints are meant to be solved for the 7 moments related to $\pi_a$,
but there are still other 6 variables related to the degree of freedom $(a,\pi_a)$;
namely the expectation value $a$ itself and its fluctuation $(\Delta a)^2$,
as well as its correlations $\Delta(a \pi_\phi)$, $\Delta(a \phi)$,
$\Delta(a \pi_\textbf{k})$, $\Delta(a v_\textbf{k})$. These are the 6 gauge
parameters of our system, and in order to reduce them just to one, it is
necessary to impose 5 gauge conditions. Since we want to interpret $a$ as time,
it is natural to request that it is a parameter with vanishing fluctuations
and correlations. Therefore, the most natural gauge choice is given by the conditions,
\begin{eqnarray}\label{gaugefixing1}
\phi_1 &=& (\Delta a)^2=0,\\
\phi_2 &=& \Delta(a \pi_\phi) =0,\\
\phi_3 &=& \Delta(a \phi) =0,\\
\phi_4 &=& \Delta(a \pi_\textbf{k})=0,\\\label{gaugefixinglast}
\phi_5 &=& \Delta(a v_\textbf{k})=0.
\end{eqnarray}
It can be shown that, once constraints and gauge conditions are imposed,
the gauge conditions $\phi_i$ commute among themselves and with
the physical variables (expectation values and all moments that do
not involve any $\pi_a$ or $a$). Therefore, they also commute with
$C_1$, so that it remains first class. In addition, the constraint
$C_2$ also remains first class but it acts trivially on the physical
variables, so it does not generate any flow on the surface defined by the
constraints and gauge conditions. This means that the Hamiltonian
$\eqref{totalham}$ with Lagrange multipliers $\mu_1=1$ and $\mu_i=0$
for $i=2,...,7$, conserves these gauge conditions throughout evolution.

Furthermore, in the Appendix B we show that taking the constraints $C_3-C_7$
and the gauge-fixing conditions \eqref{gaugefixing1}--\eqref{gaugefixinglast}
as a new whole set of constraints, their commutation
matrix that is formed by their Poisson brackets and it is necessary to
construct the Dirac brackets, is invertible in the general case.
This proves the adequacy of the chosen gauge.

In summary, we end up with the Hamiltonian $(\pi_a+{\cal H}_a)$,
where gauge conditions \eqref{gaugefixing1}--\eqref{gaugefixinglast} should be imposed.
Furthermore, $a$ plays the role of the evolution parameter and the physical variables are
the 4 expectation values $\phi$, $\pi_\phi$, $v_\textbf{k}$ and $\pi_\textbf{k}$,
together with their 4 fluctuations and the 6 correlations between them. In fact,
since these physical variables commute with $\pi_a$ we can define the object
${\cal H}_a$ \eqref{quantumhamiltonian} as the physical Hamiltonian of our system.

\subsection{Power spectra}

The power spectrum for the variable $v$ can be defined as,
\begin{equation}
P_v(k)=\frac{k^3}{2\pi^2}(\Delta v_\textbf{k})^2.\label{powerspectrumfluctuation}
\end{equation}
Nonetheless, since this quantity is not observed directly at the CMB,
it is convenient to write the power spectra of quantities related to measurable magnitudes.
On the one hand, scalar perturbations are present in temperature anisotropies on the CMB.
They are related to the comoving curvature perturbation 
\begin{equation}
\mathcal{R}:= \frac{H}{\phi'}v_\textbf{k}.
\end{equation}
Therefore, from (\ref{powerspectrumfluctuation}), the power spectrum of scalar perturbations can be written as
\begin{equation}
P_S(k) := P_\mathcal{R}(k) = \frac{H^2}{\phi'^2} P_v(k) = \frac{H^2}{\phi'^2} \frac{k^3}{2\pi^2}(\Delta v_\textbf{k})^2
= \frac{3 G}{2 \epsilon}\frac{k^3}{\pi^2 a^2}(\Delta v_\textbf{k})^2.\label{powerspectrumscalar}
\end{equation}
In the last equality we have used the relation between the slow-roll parameter $\epsilon$ and
the derivative of the scalar field $\phi$ that can be obtained from Friedmann equations. 
Tensor perturbations, on the other hand, are expressed with the quantity 
\begin{equation}
h_\textbf{k} := \frac{2\sqrt{3G}}{a} v_\textbf{k}.
\end{equation}
Due to the fact that tensor perturbations have two polarizations, the tensorial power spectrum will have a factor of 2 with respect to the power spectrum of $h_\textbf{k}$,
\begin{equation}
P_T(k):= 2P_h(k) = 2 \frac{12 G}{ a^2} P_v(k) = 12 \frac{G k^3}{\pi^2 a^2}(\Delta v_\textbf{k})^2.\label{powerspectrumtensor}
\end{equation}
These power spectra will be evaluated at late times ($|k\eta| \rightarrow 0$ or equivalently $a\rightarrow \infty$) \cite{Kin05}.
As can be seen both scalar \eqref{powerspectrumscalar} and tensorial \eqref{powerspectrumtensor} power spectra are proportional to the dimensionless ratio,
\begin{equation}\label{ratio}
{\cal P}:=\frac{G k^3 (\Delta v_\textbf{k})^2}{a^2}.
\end{equation}
Therefore, in the next section we will analyze the behavior of this object for the case of a de Sitter universe.

\section{de Sitter universe}

\subsection{Classical de Sitter background}

Since we are interested in computing quantum corrections to the de Sitter power spectra, we will first obtain the result in the approximation of QFT on classical backgrounds. This is a well-known result but, for clarity,
we think it is interesting to make it explicit in terms of this moment formalism.

The de Sitter universe is characterized by a constant value of the
Hubble factor $H$. This universe can be recovered from the model 
presented above just by imposing a constant value for
the potential $V(\phi)=H^2/(2 G)$, which fixes the energy of inflation.
The evolution of the scale factor in terms of the conformal
time $\eta$ is given by
\begin{equation}\label{desittera}
a(\eta) = \frac{-1}{H\eta}.
\end{equation}

Since in the de Sitter case the slow-roll parameter $\epsilon$ vanishes, the frequency
of the scalar sector \eqref{scalarfreq} coincides with the one of the tensorial sector \eqref{tensorfreq},
and takes the following form,
\begin{equation}
\omega_S^2 = \omega_T^2 = \frac{a''}{a} = k^2 - \frac{2}{\eta^2} = k^2 \left(1 - \frac{2}{\xi}\right),
\end{equation}
where we have introduced the dimensionless time $\xi:=-k\eta$ in order to simplify expressions below.
By replacing this form of the frequency in the equation of the fluctuation
$(\Delta v_\textbf{k})^2$ \eqref{deltav2general}, one gets
\begin{equation}
\xi^3 \frac{d^3 (\Delta v_\textbf{k})^2}{d\xi^3} + 4\xi (\xi^2-2) \frac{d (\Delta v_\textbf{k})^2}{d\xi}+8 (\Delta v_\textbf{k})^2 = 0.\label{deltav2desitter}
\end{equation}
This equation can be solved analytically, which yields the following form
of the fluctuation of the perturbative master variable,
\begin{equation}
(\Delta v_\textbf{k})^2=  \xi \left(A (J_{3/2}(\xi))^2 + B J_{3/2}(\xi)Y_{3/2}(\xi) + C (Y_{3/2}(\xi) )^2\right), \label{generalsoldesitter}
\end{equation}
where $J_{3/2}$ and $Y_{3/2}$ are the first- and second-type Bessel functions respectively, which
can be written in terms of elementary trigonometric functions,
\begin{equation}
J_{3/2}(\xi) = \sqrt{\frac{2}{\pi\xi}} \left(\frac {\sin(\xi)}{\xi} - \cos(\xi)\right),
\end{equation}
\begin{equation}
Y_{3/2}(\xi) = -\sqrt{\frac{2}{\pi \xi}} \left( \frac{\cos(\xi)}{\xi} + \sin(\xi)\right).
\end{equation}

In order to fix the integration constants $A$, $B$ and $C$, we consider the Bunch-Davies vacuum
\eqref{vpiconditions}--\eqref{bdvacuum} as
initial condition which, as explained above, corresponds to the state
with the minimum energy at the beginning of inflation ($\xi \rightarrow \infty$).
With this initial state, the evolution of the three second-order moments
is then given by
\begin{eqnarray}
(\Delta v_\textbf{k})^2 (\xi) &=&
\frac{\hbar}{2k} \frac{1+\xi^2}{\xi^2},\label{classicalms1}
\\\label{classicalms2}
(\Delta \pi_\textbf{k})^2(\xi) &=& 
\frac{k \hbar}{2}\frac{\xi^4-\xi^2+1}{\xi^4},
\\\label{classicalms3}
\Delta (v_\textbf{k} \pi_\textbf{k})(\xi) &=&  \frac{\hbar}{2\xi^3}.
\end{eqnarray}
In particular, from these expressions, it can be explicitly seen
that the uncertainty principle \eqref{uncertainty} is saturated for all $\xi$. 
For later convenience, since in the quantum-background approach
the time parameter will be the scale factor $a$, we reparametrize
the fluctuation of the perturbative master variable by using the expression \eqref{desittera},
\begin{equation}\label{deltvaqft}
(\Delta v_\textbf{k})^2 (a) = \frac{\hbar a^2 H^2}{2k^3} + \frac{\hbar}{2k}.
\end{equation}
In the late-time limit $a \rightarrow \infty$, the second term above is then suppressed
and the standard result for the power spectrum \eqref{ratio} is obtained,
\begin{equation}\label{psqftdesitter}
{\cal P}=\frac{G k^3}{a^2}(\Delta v_\textbf{k})^2=\frac{\widetilde{H}^2}{2},
\end{equation}
where we have defined the dimensionless Hubble parameter
$\widetilde{H}:=l_p H$, with the reduced Planck length $l_p:=\sqrt{G\hbar}=(4\pi\hbar G_N/3)^{1/2}$.
One of the main properties of this power spectrum is
its scale invariance, since it does not depend on the wave number $k$.

\subsection{Quantum de Sitter background}

The quantum-gravity corrections for the power spectrum will be now analyzed, where the dynamics of the variables is governed by the physical Hamiltonian  ${\cal H}_a$, as defined in \eqref{quantumhamiltonian}, with gauge conditions \eqref{gaugefixing1}--\eqref{gaugefixinglast}. The amount of physical variables is now 14: the 4 expectation values ($\phi$, $\pi_\phi$, $v_\textbf{k}$, $\pi_\textbf{k}$), their 4 fluctuations and the 6 correlations between each couple. The evolution is then given by the solutions of the system of 14 differential equations of the form
$dX/da=\{X,\mathcal{H}_a\}$. In the general case the equations turn out to be quite complicate but,
as will be shown below, in the de Sitter case it will be possible to find an approximate analytic solution.

As in the previous case we impose the de Sitter conditions
$V=H^2/(2 G)$ and $\omega^2=k^2-2a^2H^2$. In addition,
for the expectation values $v_{\bf k}$, $\pi_{\bf k}$ and $\pi_\phi$,
we will assume that initially they take the same value as in
the previous section, that is, $v_{\bf k}=0$, $\pi_{\bf k}=0$ and $\pi_\phi=0$.
The first two values come from the stationary-state analysis discussed in Section \ref{moments_classical_back},
whereas the last one is motivated from the classical behavior of this variable.
It turns out that our quantum equations of motion preserve those values.
That is, once replaced all the commented conditions, one gets that
at this order the expectation values $\phi$, $\pi_\phi$, $v_{\bf k}$,
and $\pi_{\bf k}$ are constants of motion. Therefore, concerning these variables,
the stationary point and the classical de Sitter behavior are respected.
This is a nontrivial result, since the mentioned equations of motion
might well have been modified by different moments in such a way that,
for instance, ($v_{\bf k}=0$, $\pi_{\bf k}=0$) was not a stationary point
of the system anymore.

In summary, once imposed the above conditions, out of the 14 equations of
motion, 5 are trivial. The expectation values of the field $\phi$ and of the perturbative variables
$v_{\bf k}$, as well as of their momenta $\pi_\phi$ and $\pi_{\bf k}$, are constants of motion:
\begin{eqnarray}
 \frac{d}{da} \phi&=&0,\\
 \frac{d}{da} \pi_\phi&=&0,\\\label{v0constant}
  \frac{d}{da} v_{\bf k}&=&0,\\
   \frac{d}{da} \pi_{\bf k}&=&0.
 \end{eqnarray}
Furthermore, the fluctuation of the momentum of the
field is also conserved through evolution,
\begin{equation}
    \frac{d}{da} (\Delta \pi_\phi)^2=0.
\end{equation}
Finally, the other 9 nontrivial equations take the following form:
\begin{eqnarray}\label{quantumeq1}
G{\cal H}_a \frac{d}{da}((\Delta \phi)^2)&=&\frac{2}{a^2} \Delta(\phi\pi_\phi)\\
G{\cal H}_a \frac{d}{da}(\Delta(\phi\pi_\phi))&=&\frac{1}{a^2}(\Delta \pi_\phi)^2,
   \\
G{\cal H}_a \frac{d}{da}(\Delta(\phi v_{\bf k}))&=&\Delta(\phi \pi_{\bf k})
+\frac{1}{a^2}\Delta(\pi_\phi v_{\bf k}),
\\G{\cal H}_a
\frac{d}{da}(\Delta(\phi \pi_{\bf k}))&=&
\Delta(\phi v_{\bf k}) \left(2 a^2 H^2-k^2\right)+\frac{1}{a^2}\Delta(\pi_\phi \pi_{\bf k}),
\\
G{\cal H}_a \frac{d}{da}(\Delta(\pi_\phi v_{\bf k}) )&=&\Delta(\pi_\phi \pi_{\bf k} ),
\\\label{quantumeqlast}
G{\cal H}_a \frac{d}{da}(\Delta(\pi_\phi \pi_{\bf k}))&=&- \Delta(\pi_\phi v_{\bf k}) \left(k^2-2 a^2 H^2\right),
\\\label{quantumms1}
G{\cal H}_a \frac{d}{da}((\Delta v_{\bf k})^2)&=&2  \Delta(v_{\bf k}\pi_{\bf k}),\\\label{quantumms2}
G{\cal H}_a\frac{d}{da}((\Delta \pi_{\bf k})^2)&=&-2  \Delta(v_{\bf k}\pi_{\bf k}) \left(k^2-2 a^2 H^2\right),
   \\\label{quantumms3}
G{\cal H}_a \frac{d}{da}(\Delta(v_{\bf k}\pi_{\bf k}))&=&(\Delta v_{\bf k})^2 \left(2 a^2 H^2-k^2\right)+(\Delta \pi_{\bf k})^2,
\end{eqnarray}
with the physical Hamiltonian given by,
\begin{equation}\label{physicalham}
{\cal H}_a= \frac{1}{\sqrt{G}} \left[\frac{a^4}{G} H^2+(\Delta v_{\bf k})^2
\left(k^2-2 a^2 H^2\right)+(\Delta \pi_{\bf k})^2
+\frac{1}{a^2}(\Delta \pi_\phi)^2\right]^{1/2}.
\end{equation}
This is a highly coupled and nonlinear system of differential equations, in particular due to the form of this
physical Hamiltonian, and analytical solutions
are extremely difficult to get.
Looking at the equations, we note that it would be natural to define a time $\tau$, in such a way that $G{\cal H}_a d\tau= da$,
so that the left-hand side of all equations would be just the derivative
of the corresponding variable with respect to this time.

The limit of QFT on classical backgrounds is easily recovered from the system above just
by dropping all moments that involve any background object. In particular, equations \eqref{quantumeq1}--\eqref{quantumeqlast} are trivially obeyed in that limit. Furthermore, to obtain that limit,
the physical Hamiltonian should also be taken as a classical variable
(since it is the momentum $\pi_a$), so one should drop all moments
from its definition, which leads to ${\cal H}_a=a^2 H/G$. In this case
the natural time $\tau$ is the standard conformal time $\tau=\eta=-1/(H a)$ and equations
\eqref{quantumms1}--\eqref{quantumms3} for the fluctuations of the perturbative variables
reduce to their form \eqref{qfteq3}--\eqref{qfteq5}.

It is interesting to note that, since
the fluctuation of the momentum of the field $(\Delta \pi_\phi)^2$, which
appears inside the Hamiltonian, is a constant of motion, equations \eqref{quantumms1}--\eqref{quantumms3}
for the moments related to the perturbative degree of freedom $(v_{\bf k},\pi_{\bf k})$ decouple from
the rest. In particular, it is possible to write the generalization of equation \eqref{deltav2desitter} for the fluctuation of the perturbative master variable as follows,
\begin{equation}\label{quantumeqdeltav}
{\cal H}_a^2 \frac{d^3(\Delta v_{\bf k})^2}{da^3} +3
   {\cal H}_a \frac{{d\cal H}_a}{da} \frac{d^2(\Delta v_{\bf k})^2}{da^2}+
\left[4 \frac{\omega^2}{G^2}+{\cal H}_a
   \frac{{d^2\cal H}_a}{da^2} +\left( \frac{{d\cal H}_a}{da} \right)^2\right]\frac{d(\Delta v_{\bf k})^2}{da} -\frac{8}{G^2} a H^2  (\Delta v_{\bf k})^2=0.
\end{equation}
Nevertheless, contrary to the QFT version of this equation \eqref{deltav2desitter},
this full-quantum equation cannot be solved independently from the rest, since
it is still coupled to the fluctuation of the momentum $(\Delta \pi_{\bf k})^2$
and their correlation $\Delta(v_{\bf k}\pi_{\bf k})$ through the Hamiltonian
${\cal H}_a$.
Finally,
if the dependence of the Hamiltonian on $a$ was known, one could write that equation
in terms of the time $\tau$ in a very compact way,
\begin{equation}
 \frac{d^3}{d\tau^3}(\Delta v_{\bf k})^2+4\omega\frac{d}{d\tau}(\omega(\Delta v_{\bf k})^2)=0.
\end{equation}
Although this might seem a simple equation to deal with, all the complexity is inside
the frequency $\omega=\omega(\tau)$ and analytic solutions are very hard to find.

\subsubsection{Approximate analytical solution for the power spectrum}

Since one expects to find small changes to the result of the previous section,
a first approximation to obtain the evolution of the fluctuation $(\Delta v_{\bf k})^2$
would be to assume the behavior derived in the approximation of QFT on classical
backgrounds to compute the evolution of the physical Hamiltonian ${\cal H}_a(a)$.
Plugging this form into \eqref{quantumeqdeltav} will decouple this equation
from the rest and one can then solve it to obtain the approximate dynamics of
the fluctuation $(\Delta v_{\bf k})^2$.

Therefore, replacing the form of the moments \eqref{classicalms1}-\eqref{classicalms3}
in the definition of the Hamiltonian \eqref{physicalham} one gets,
\begin{equation}
{\cal H}_a=
\frac{a^2 H}{G}\sqrt{1
+\frac{\widetilde{H}^2}{2 k^3} \left(2 \xi ^4-2 \xi^2-1\right)
+\frac{\widetilde{H}^4}{k^6} \sigma^2 \xi^6}.\label{physicalhamiltonapprox}
\end{equation}
In order to absorb some of the fundamental constants in this expression
we have made use of the dimensionless Hubble parameter $\widetilde{H}:=l_p H$ and
the time parameter $\xi:= k/(aH)$, as already defined above;
whereas the constant fluctuation of the momentum $(\Delta \pi_\phi)^2$ has been described by the quantity
$\sigma$ as follows: $(\Delta \pi_\phi)^2=:G\hbar^2 \sigma^2$. In this way,
it is shown very explicitly that $\widetilde{H}$ is the parameter that measures
the strength of quantum-gravity effects. Since this parameter is known to be
quite small \cite{Ade:2015xua}, one can perform a power expansion of the square root above
and write the Hamiltonian in the following way,
\begin{eqnarray}\label{expham}
{\cal H}_a&=&\frac{a^2 H}{G}\sum_{n=0}^\infty\left( \frac{\widetilde H^2}{k^3}\right)^n f_n(\xi)
\\\nonumber
&= &
\frac{a^2 H}{G}\left[ 1 +\frac{\widetilde{H}^2}{4 k^3} \left(2 \xi ^4-2 \xi ^2-1\right)
 -\frac{\widetilde{H}^4}{32 k^6} \left[4 \xi ^8-8 \xi ^6 \left(2 \sigma ^2+1\right)+4 \xi^2+1\right]+{\cal O}\left(\frac{\widetilde H^6}{k^9}\right)\right],
\end{eqnarray}
where the series only contains even
powers of $\widetilde{H}/k^{3/2}$. It is remarkable that the quantity $\sigma$
only enters at order $\widetilde H^4$, which means that the actual state of
the fluctuations of the matter field is not the most dominant part among
the quantum-gravity corrections.

One can then assume a similar power expansion
for the fluctuation $(\Delta v_{\bf k})^2$, around its value in
the QFT approximation \eqref{classicalms1}, as follows
\begin{equation}\label{expv}
(\Delta v_{\bf k})^2=\frac{\hbar}{k}\left[\frac{\xi^2+1}{2 \xi ^2} +\sum_{n=1}^\infty\left( \frac{\widetilde H^2}{k^3}\right)^n\delta_n(\xi) \right],
\end{equation}
with unknown functions $\delta_n(\xi)$.
Replacing the last two expressions \eqref{expham}--\eqref{expv} in equation \eqref{quantumeqdeltav},
and taking into account the definition of the dimensionless time $\xi=k/(a H)$,
one can solve for the different functions $\delta_n(\xi)$ iteratively at each order
in $\widetilde H^2$.

The power spectrum can be computed just by replacing the above form of $(\Delta v_{\bf k})^2$
in its definition \eqref{ratio} and it takes the form 
\begin{equation}\label{psexpansion}
{\cal P}=\frac{G k^3}{ a^2}(\Delta v_{\bf k})^2=\frac{\widetilde{H}^2}{2}\left[1+\xi^2+2 \sum_{n=1}^\infty\left(\frac{\widetilde H^2}{k^3}\right)^n \xi^2\delta_n(\xi)\right].
\end{equation}
In order to obtain its exact numerical value, the different $\delta_n$ functions must be obtained
as indicated above. Nonetheless, from this expression
one can already see the explicit dependence of the power spectrum on the different parameters, like
the Hubble factor $H$ and the wave number $k$. In particular we see that the quantum-gravity corrections
are given by a power series in the parameter $\widetilde H^2/k^3$, which makes it scale dependent
and more relevant for large scales, which are related with small values of the mode
number $k$. We would like to emphasize that the very same form for the first term of this series
has been obtained in several other studies with very different approximation methods,
see for instance \cite{brizuelakieferkramer1, kieferkramer, KTV13}.

In order to know whether these correction terms imply an enhancement or a suppression of power, we need to study
the behavior of the $\delta_n$ functions in the asymptotic limit of super-Hubble scales.
Since the expansion parameter is very small, we will just focus on the first correction term given by $\delta_1$,
which obeys the following equation,
\begin{equation}
\xi ^5 \delta_1'''(\xi )+4 \xi ^3 (\xi^2 -2) \delta_1'(\xi )+8 \xi ^2 \delta_1(\xi )+4 \xi ^2+6=0.
\end{equation}
The analytical solution of this equation is obtained and the constants are fixed by requesting that the solution
is not oscillating at the beginning of inflation ($\xi\rightarrow \infty$). In this way, we find
the following form for the first corrective term to the evolution of the fluctuation of
the perturbative variable $(\Delta v_{\bf k})^2$:
\begin{equation}
\delta_1\!=\!\frac{1}{12\xi^2}\left\{11+
4 \text{Ci}(2 \xi ) \left[\left(\xi ^2-1\right) \cos (2 \xi )-2 \xi  \sin (2
\xi )\right]\!-\!2\left[\left(\xi ^2-1\right) \sin (2 \xi )+2  \xi  \cos (2 \xi ) \right] (\pi -2 \text{Si}(2 \xi ))\right\}\!,\label{delta1}
\end{equation}
$\text{Si}$ and $\text{Ci}$ being respectively the sine and cosine integral functions.
The asymptotic limit for super-Hubble scales ($\xi\rightarrow 0$) for the coefficient that
appears multiplying the correction term of order $\widetilde H^2/k^3$ in \eqref{psexpansion}
is given by,
\begin{equation}\label{delta1sol}
2 \xi^2 \delta_1=\frac{1}{6} (11-4\gamma_E-4 \ln 2\xi)+{\cal O}(\xi),
\end{equation}
$\gamma_E$ being the Euler-Mascheroni constant. This expression still contains certain time dependence
in the last logarithmic term, which slowly diverges. In order to give an estimate, it
is usual to evaluate this expression at horizon crossing ($\xi=1$), which
gives the following numerical value,
\begin{equation}
 2 \xi^2 \delta_1\approx 0.986.
\end{equation}
Hence, up to the considered order, the power spectrum takes the form
\begin{equation}
{\cal P}\approx\frac{\widetilde{H}^2}{2}\left[1+\frac{\widetilde{H}^2}{k^3}\right]+{\cal O}(\widetilde H^4),\label{spectrumanalytical}
\end{equation}
where the above numerical value has been rounded up to one.
This is one of the main results of this paper and implies an enhancement of the power spectrum
for large scales. In fact, the numerical value of the corrections obtained in this analysis
is in agreement with the one computed in \cite{brizuelakieferkramer1},
where a very different semiclassical approximation scheme was used.
This is certainly an important indication of the robustness of the presented result.

\subsubsection{Numerical analysis of the system}

In order to show the validity of the approximate analytical solution obtained above, in this
subsection we will solve numerically the full system of quantum equations \eqref{quantumeq1}--\eqref{quantumms3}.
In fact, as already commented above, since the fluctuation of the momentum of the field $(\Delta \pi_\phi)^2=G \hbar^2\sigma^2$
is a constant of motion, equations \eqref{quantumms1}--\eqref{quantumms3} for the moments related to the
perturbative degree of freedom $(v_{\bf k},\pi_{\bf k})$ decouple from the rest. Since, in order to compute
the power spectrum, we are interested in the behavior of the fluctuation $(\Delta v_{\bf k})^2$, it is
enough to consider those three equations.
For the rest of this section we will choose units with $\hbar=1$ and $G=1$.

The main objective is to compare the correction of the power spectrum for both the approximate analytical
and numerical approaches. We thus define the relative correction between the quantum and classical background
results as,
\begin{equation}
\beta(a) = \frac{(\Delta v_\mathbf{k})^2_{QG}(a)-(\Delta v_\mathbf{k})^2_{QFT}(a)}{(\Delta v_\mathbf{k})^2_{QFT}(a)},
\end{equation}
where $(\Delta v_\mathbf{k})^2_{QG}$ stands for the solution of the full quantum equations
\eqref{quantumms1}--\eqref{quantumms3}
(obtained either analytically as in \eqref{expv}--\eqref{delta1sol} or numerically in this subsection) and $(\Delta v_\mathbf{k})^2_{QFT}$ corresponds to the solution \eqref{deltvaqft} of QFT on classical backgrounds.
In fact, within the decomposition \eqref{expv} performed in the previous subsection,
this object $\beta$ is related to the $\delta_n$ functions. More precisely
$\beta$ is equal to the third-term inside the square brackets of equation \eqref{psexpansion},
that provides the relative correction to the power spectrum, and its explicit form
within the approximation considered there is given by,
\begin{equation}\label{approxbeta}
 \beta\approx \frac{2\widetilde H^2}{k^3}\xi^2\delta_1,
\end{equation}
with $\delta_1$ given by \eqref{delta1}. In the following we will compare this result with
the numerical solution of the full system.

One problem in the numerical approach is that the system \eqref{quantumms1}--\eqref{quantumms3}
is singular at $a=0$. Thus that point can not be used to impose the initial conditions and
there is no other preferred choice for the initial value of the scale factor $a_0>0$.
We have therefore used two different methods to impose the initial state numerically.
On the one hand, in the first method we have assumed the Bunch-Davies vacuum
\eqref{vpiconditions}--\eqref{bdvacuum} at an initial value of the scale factor $a_0$.
On the other hand,
in the second method we have used the analytical expression for the quantum variables obtained above \eqref{expv}--\eqref{delta1sol}
(where the Bunch-Davies vacuum has been exactly implemented for $a_0 = 0$) as initial conditions for the
numerical simulation at a finite value of $a_0$.

\begin{figure}[t]
\centering
\includegraphics[scale=0.22]{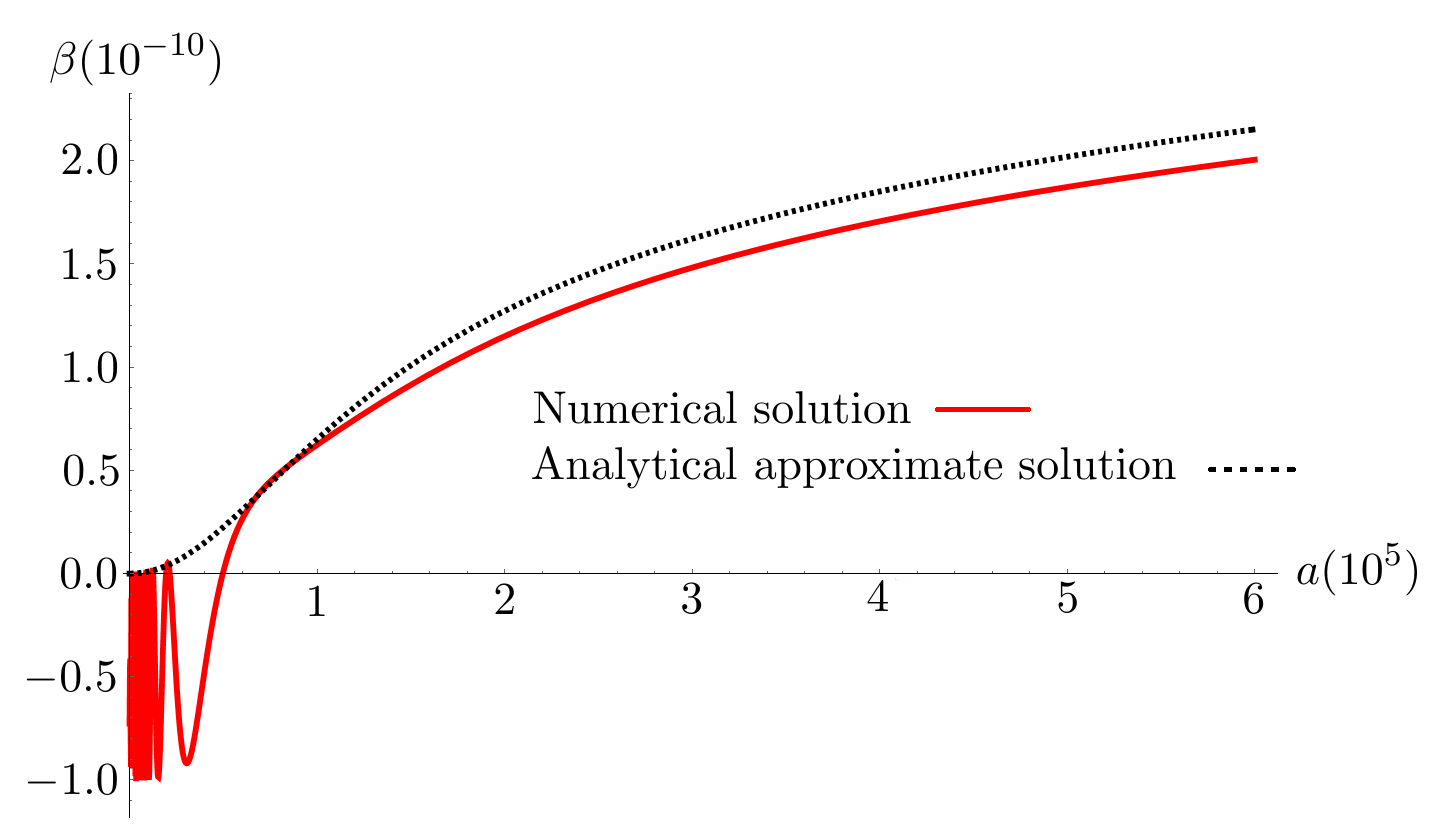}
\includegraphics[scale=0.22]{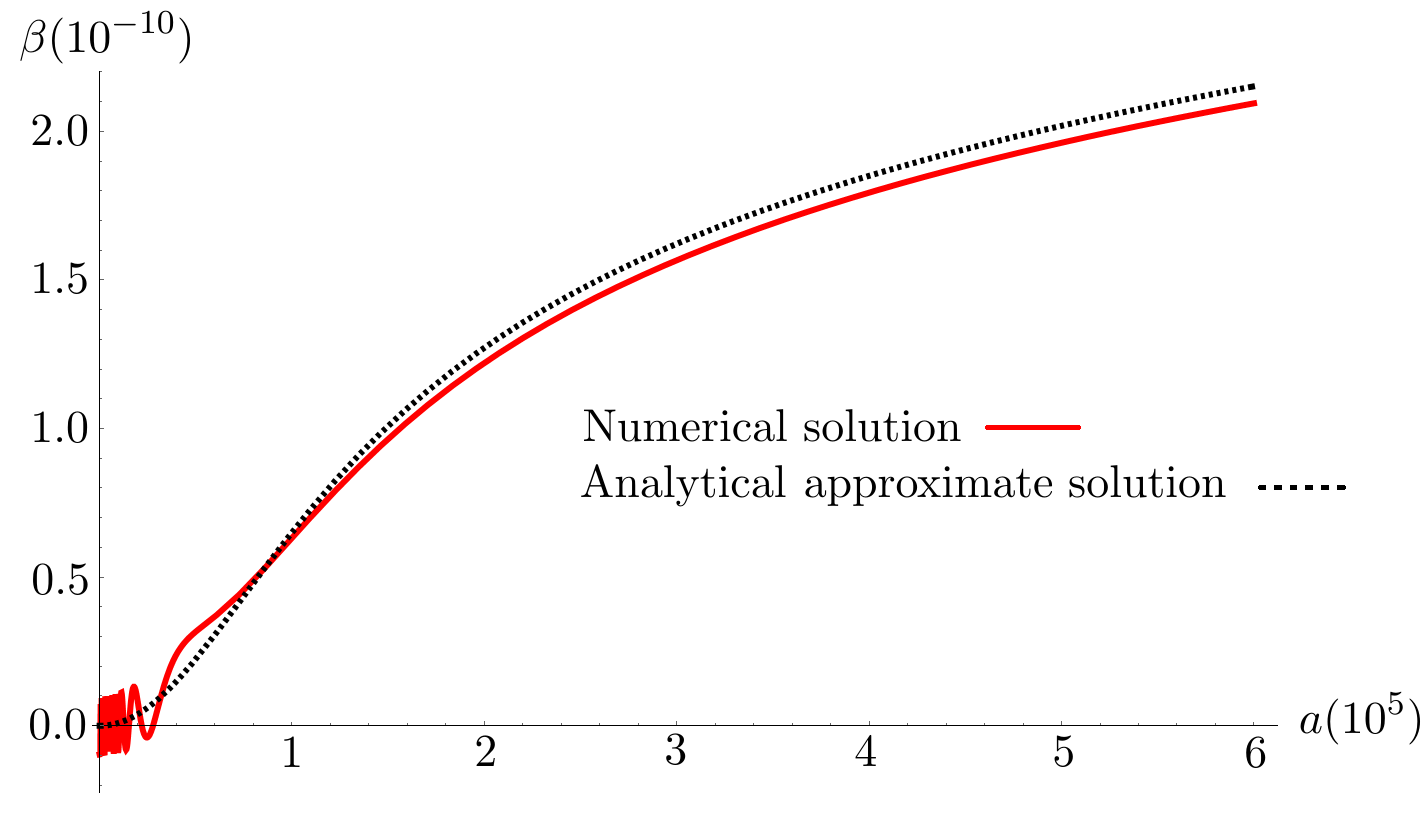}
\caption{The relative correction of the power spectrum $\beta$, as computed by the numerical solution of the system \eqref{quantumms1}--\eqref{quantumms3} and the approximate analytical solution \eqref{expv}--\eqref{delta1sol}, for
$H = 10^{-5}$, $\sigma = 1/10$, $k=1$ and an initial value of the scale factor $a_0 = 10^{-15}$.
The numerical evolution shown in the plot on the left-hand side corresponds to the first method (an initial Bunch-Davies vacuum), whereas the numerical evolution shown on the right-hand side corresponds to the second method (initial state given by the approximate analytical solution). In both cases it can be observed that for large scales the numerical evolution of $\beta$ is
qualitatively the same as the evolution given by the approximate solution. In fact, their difference is small and
approximately constant in time.
} \label{relativeerror} 
\end{figure}

In summary, for both methods, although the evolution is sensitive to the initial value of $a_0$, we observe
that for $a_0 \ll 1$ the solutions are robust and have the same asymptotic behavior for large scales.
In particular, we have performed tests with $a_0 = 10^{-10}$, $a_0 = 10^{-15}$ and $a_0 = 10^{-20}$,
and the evolution of all variables remains invariant for large values of the scale factor.
In addition, for both methods, the value of $\beta$ for large scales is approximately the same.
Furthermore, for small values of the fluctuation of the field $\sigma\lessapprox 1$,
the asymptotic value of the relative correction $\beta$ coincides with the one obtained with the
approximate solution \eqref{approxbeta} with high accuracy,
which validates the analytical result presented in the previous subsection.

More precisely, in Fig. \ref{relativeerror} the evolution of the relative correction $\beta$ as computed with the approximate analytical
and with the numerical solution is shown. The plot on the left corresponds to the first method of
imposing the initial conditions, and the plot on the right to the second one. In both cases 
$a_0 = 10^{-15}$, $k=1$, $\sigma=1/10$, and $H = 10^{-5}$ have been chosen.
This value of the Hubble factor is in accordance with the experimental data \cite{Ade:2015xua}.
In the numerical evolutions two different regimes can be clearly identified. First, for small values of $a$,
the function $\beta$ is oscillatory. The difference between the first and the second method is more relevant
in this regime, as the amplitude of the oscillations and their average value is smaller in the latter case.
Then, at certain point, $\beta$ stops oscillating and it becomes a monotonically
increasing function. In this regime both methods provide very similar results,
with the same qualitative evolution as the one predicted by the approximate analytical solution.
In this second regime, it is possible to see that the numerical value of $\beta$ differs slightly
from the one obtained analytically, but this difference is small and it remains approximately constant in time.
In the particular case shown in Fig. \ref{relativeerror}, at the beginning of this second regime
(at around $a \approx 2\cdot10^5$) the relative difference between the numerically computed and analytically
approximated $\beta$ is around $10\%$ in the left plot and $5\%$ in the right plot.

\begin{figure}[t]
\centering
\includegraphics[scale=0.18]{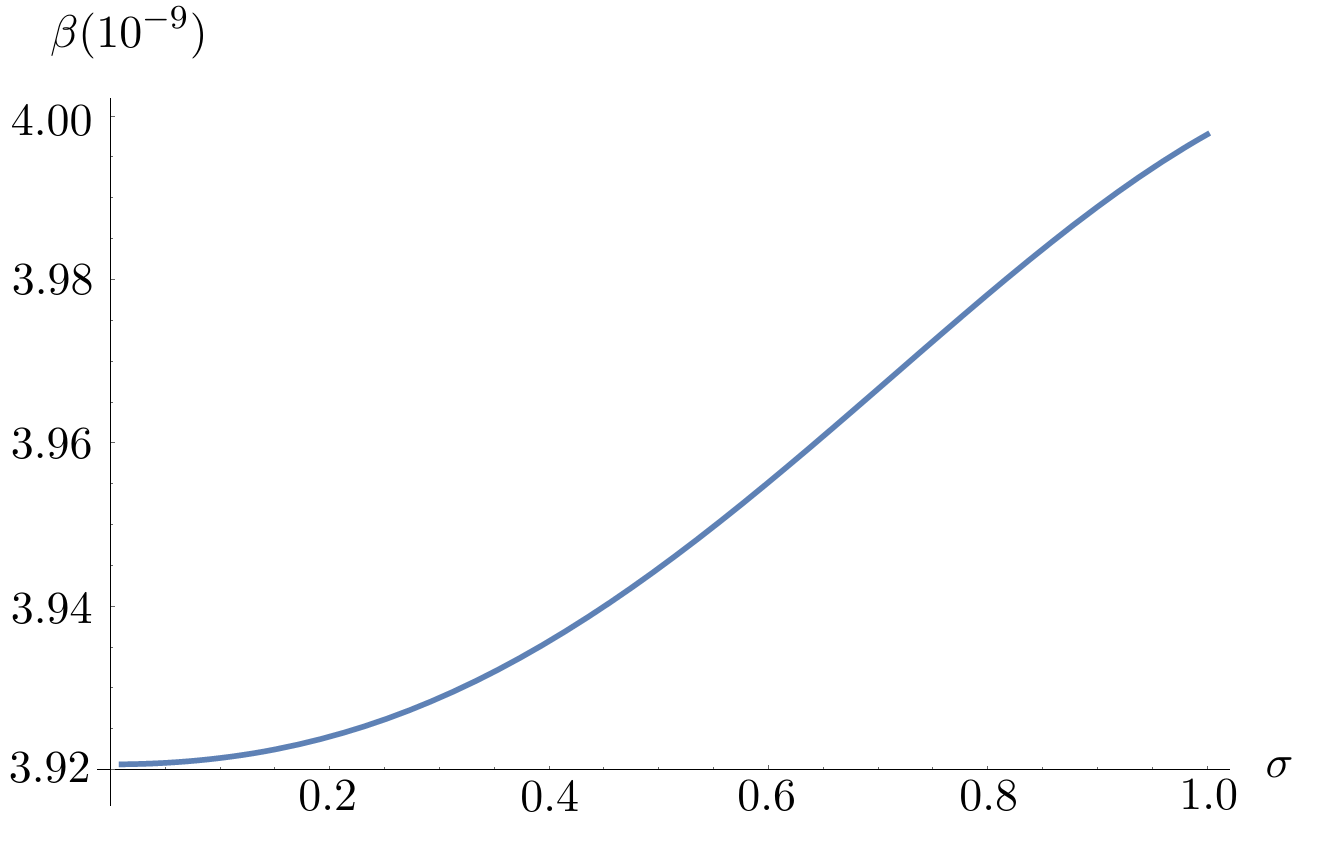}
\includegraphics[scale=0.18]{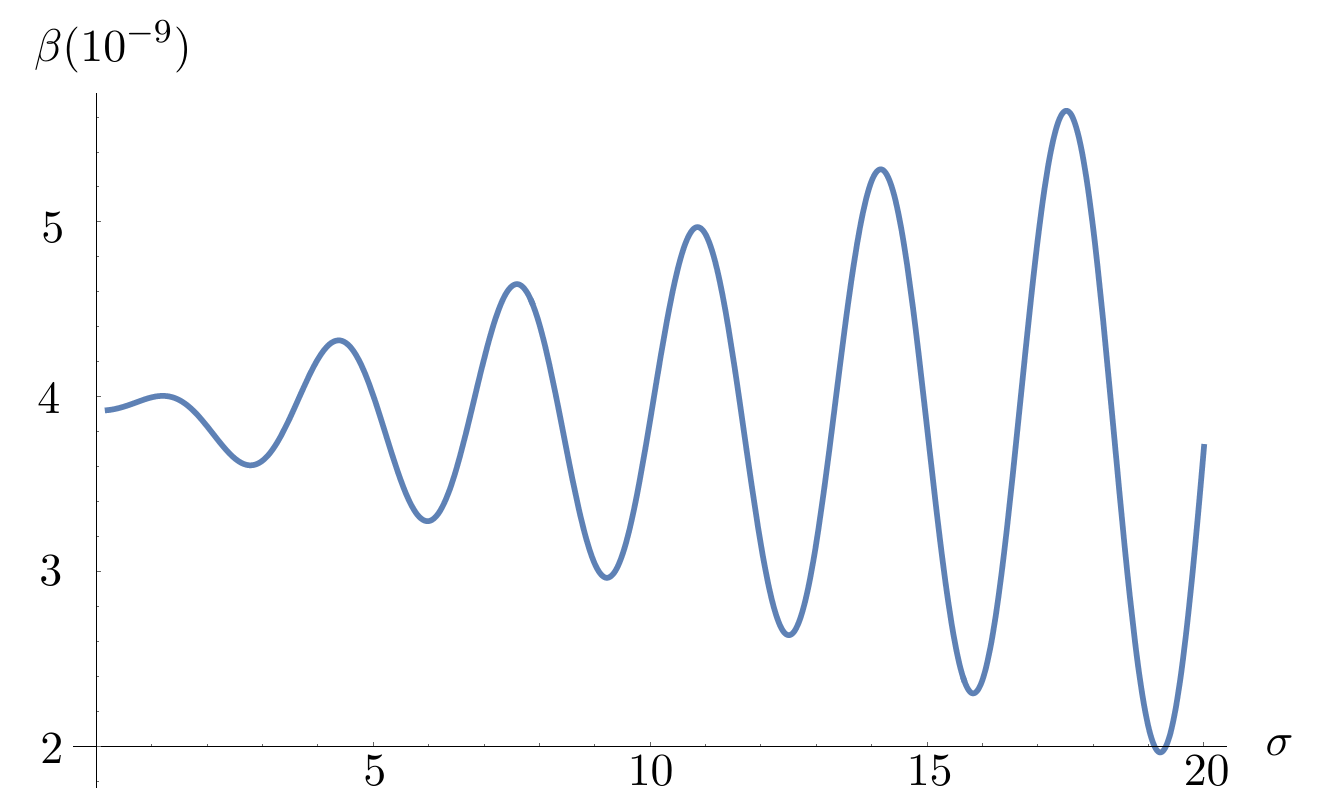}
\caption{The relative correction of the power spectrum $\beta$ as a function of $\sigma$, for $H = 10^{-5}$ and $k=1$,
evaluated at $a=10^{30}$. Each plot shows different ranges of values for $\sigma$.}\label{sigmaplot}
\end{figure}

Finally, in the analytical discussion it has been observed that the dependence of the
quantum-gravity corrections on $\sigma$, which describes the quantum state of the
scalar field, are of order $\widetilde H^4$. Therefore, as long as $\sigma$ is
not large, its effects should be negligible. In order to see if the numerical solution
validates this conclusion, in Fig. \ref{sigmaplot} the relative correction $\beta$ evaluated at a
large value of $a=10^{30}$ is plotted as a function of $\sigma$. It can be seen
that $\beta(\sigma)$ follows an oscillatory behavior, with average at $\beta(\sigma = 0)$,
and a growing amplitude for larger values of $\sigma$.
In particular, for $\sigma \simeq 10$,
as shown in the plot on the right, the amplitude of $\beta$ is of the same order of magnitude
as $\beta$ itself, and hence not negligible.
However, for small $\sigma$, the change in $\beta$ due to different values of $\sigma$
is one or two orders of magnitude smaller than $\beta$ itself.  
Therefore, the conclusions derived with the analytical method are correct and, as long
as the fluctuation of the momentum of the scalar field $\sigma$ is not very large,
its actual value is not relevant to compute the quantum-gravity corrections of the power spectrum.

\section{Conclusions}

In this work we have analyzed the behavior of the power spectra for both scalar and tensorial
inflationary perturbations in two different quantization schemes. To that end,
we have made use of
a formalism that is based on a decomposition of the corresponding wave function in terms of its moments.
On the one hand, the usual approximation of QFT on fixed backgrounds has been considered.
In this case, the background degrees of freedom are treated classically and the wave function that describes
the perturbative sector obeys a Schr\"odinger equation. The effective Hamiltonian that provides
the evolution of the different moments is quadratic and thus different orders in moments decouple.
In particular the power spectrum is related to the fluctuation of the master perturbative variable $(\Delta v_{\bf k})^2$,
which is a second-order moment. Therefore, in this case a cutoff is not needed and the exact study
of the dynamics at second order
is enough to obtain the well-known result for the scale-invariant power spectrum
for the de Sitter universe \eqref{psqftdesitter}.

On the other hand, the Wheeler-DeWitt quantization has been implemented, where the background
degrees of freedom have also been quantized. As opposed to the previous case, the main (Wheeler-DeWitt) equation
is not an evolution equation, but rather a constraint equation. The information on the Wheeler-DeWitt equation
is translated to an infinite set of constraint equations obeyed by the moments. In addition, the
corresponding effective Hamiltonian is not quadratic in all the variables and different orders
in moments do couple. Hence, for this scheme, it is necessary to apply a truncation. In our case
we have truncated the system at second order in moments, which is assumed to be a good approximation
for peaked semiclassical states.
Once the truncation has been performed, one gets a finite system of (first-class) constraint
equations \eqref{quantumconstraint1}--\eqref{quantumconstraint7}.
We have then considered appropriate gauge-fixing conditions \eqref{gaugefixing1}--\eqref{gaugefixinglast},
so that the scale
factor plays the role of time, and hence (minus) its conjugate momentum is the physical
Hamiltonian ${\cal H}_a$ \eqref{quantumhamiltonian}. The evolution of this system has then been analyzed in detail
for the case of a de Sitter universe. In particular, we have been able to obtain
an analytical approximate solution for the evolution of the fluctuation of the master perturbative
variable $(\Delta v_{\bf k})^2$, given by \eqref{expv} and \eqref{delta1}, which has been validated
by the subsequent numerical analysis.

In order to construct such approximate solution, we have considered the behavior given
by the approximation of QFT on fixed backgrounds for the moments that appear inside
the physical Hamiltonian. In this way, the third-order equation for the fluctuation
of the perturbative variable \eqref{quantumeqdeltav} decouples and can be solved
separately from the rest. At this point, it has been observed that this approximate physical
Hamiltonian (as well as the fluctuation of the perturbative variable) can be expanded
as a power series on the parameter $l_p^2 H^2/k^{3}$. As expected, the order zero leads to the
well-known result of QFT on classical backgrounds analyzed above, whereas the first order
provides the main quantum-gravity corrections.
After computing analytically
the solution of the system, it has been derived that the sign of the corrections
is positive, which implies an enhancement of power. Equation \eqref{spectrumanalytical} provides
the form of the power spectrum, and it is the main result of the paper.
It can be seen that the strength of the quantum-gravity corrections
is tuned by the square of the dimensionless Hubble parameter $(l_p H)^2$ and,
more importantly, that these corrections break the scale-invariance due to the
$k^{-3}$ dependence, which makes them more relevant for large scales (small $k$).
Remarkably, within this Wheeler-DeWitt quantization scheme, identical results have
been obtained by making use of very different semiclassical approximation techniques
\cite{brizuelakieferkramer1, KTV13}.

In the last subsection, by considering the numerical implementation of the full system
\eqref{quantumeq1}--\eqref{quantumms3}, we have verified
the validity of the approximate solution. In particular it has been shown that, concerning
the fluctuation of the perturbative variable $(\Delta v_{\bf k})^2$,
for small values of the scale factor $a$ the approximate and numerical solutions have different behavior,
as the former is constant through evolution whereas the latter one is oscillating.
This issue might not be something fundamental, but rather related to the chosen initial conditions.
Nonetheless, for large scales the numerical and approximate evolutions become qualitatively
identical and very close in value. Since the relevant value of the power spectrum,
that serves as a seed for the anisotropies measured at the CMB, is obtained at super-Hubble scales,
this numerical approach shows the validity of the analytical solution discussed above.

Finally, another remarkable property that is derived from the approximate solution is that
the dependence of the power spectrum on the actual state of the background matter degrees of freedom
(the inflaton field $\phi$) is of the order of $(l_p^2 H^2\sigma)^2$, $\sigma^2$ being the dimensionless
fluctuation of the momentum of the inflaton $(\Delta \pi_\phi)^2$, which is a constant of motion.
This means that, as long as the value of $\sigma$ is not too large, the result is expected not to be
much affected by the state of the matter. In fact, the explicit dependence of the power spectrum on
$\sigma$ has been studied within the numerical setting and, as can be seen in Fig. \ref{sigmaplot},
this expectation has been confirmed.

\appendix

\section{Poisson brackets between quantum constraints}

In this appendix we provide the Poisson brackets between the constraints defined
in (\ref{quantumconstraint1})-(\ref{quantumconstraint7}) up to linear order in moments,
or order $\hbar$. It turns out that all the brackets are a linear combination of constraints,
which proves that all of them are first class. In principle, with these 7 constraints,
49 different Poisson brackets can be constructed $\{C_I,C_J\}$.
However, due to the antisymmetry, 7 are identically zero, $\{C_I,C_I\}=0$, and from the other 42,
they are related by pairs $\{C_I,C_J\} = -\{C_J,C_I\}$. Consequently, there are 21
independent nontrivial brackets, which take the following form:
{\allowdisplaybreaks
\begin{eqnarray*}
\{C,C_{\pi_a}\} &=& \frac{6}{a^4}\left(\pi_\phi^2+4a^6V(\phi)\right)C_a-\frac{4}{a^3}\pi_\phi C_{\pi_\phi}+8a^3V'(\phi)C_\phi, \\
\{C,C_a\}&=&2 G C_{\pi_a},\\
\{C,C_{\pi_\phi}\} &=& 8a^3V'(\phi)C_a + 2a^4V''(\phi)C_{\pi_\phi},\\
\{C,C_{\phi}\}&=&\frac{4\pi_\phi}{a^3}C_a - \frac{2}{a^2}C_{\pi_\phi},\\
\{C,C_{\pi_{\bf k}}\}&=&2\omega^2 C_{v_{\bf k}},\\
\{C,C_{v_{\bf k}}\} &=& -2C_{\pi_{\bf k}},\\
\{C_{\pi_a},C_a\} &=&4 G \pi_a C_{\pi_a} + \frac{4}{a^3}\left(\pi_\phi^2-4a^6V(\phi)\right)C_a + \frac{2\pi_\phi}{a^2}C_{\pi_\phi}-2a^4V'(\phi)C_\phi-2\pi_\mathbf{k} C_{\pi_{\bf k}}-2\omega^2 v_\mathbf{k} C_{v_{\bf k}},\\
\{C_{\pi_a},C_{\pi_\phi}\} &=& 2a^4 V'(\phi) C_{\pi_a} + \frac{2}{a^3}\left(\pi_\phi^2-4a^6V(\phi)\right) C_{\pi_\phi},
\\
\{C_{\pi_a},C_{\phi}\} &=& - \frac{2\pi_\phi}{a^2}C_{\pi_a}+\frac{2}{a^3}\left(\pi_\phi^2-4a^6V(\phi)\right)C_\phi, \\
\{C_{\pi_a},C_{\pi_{\bf k}}\}&=&2\omega^2 v_\mathbf{k} C_{\pi_a}+\frac{2}{a^3}\left(\pi_\phi^2-4a^6V(\phi)\right)C_{\pi_{\bf k}},\\
\{C_{\pi_a},C_{v_{\bf k}}\}&=&-2\pi_\mathbf{k} C_{\pi_a}+\frac{2}{a^3}\left(\pi_\phi^2-4a^6V(\phi)\right)C_{ v_{\bf k}},
\\
\{C_a,C_{\pi_\phi}\}&=&2a^4V'(\phi)C_a - 2 G \pi_aC_{\pi_\phi},\\
\{C_a,C_{\phi}\}&=&-\frac{2\pi_\phi}{a^2} C_a - 2 G\pi_aC_{\phi},\\
\{C_a,C_{\pi_{\bf k}}\}&=&2\omega^2 v_\mathbf{k} C_a - 2 G \pi_a C_{\pi_{\bf k}},\\
\{C_a,C_{v_{\bf k}}\}&=&-2\pi_\mathbf{k} C_a - 2 G \pi_aC_{v_{\bf k}},\\
\{C_{\pi_\phi},C_\phi\} &=& 2 G \pi_a C_{\pi_a} + \frac{2}{a^3}\left(\pi_\phi^2-4a^6V(\phi)\right)C_a + \frac{4\pi_\phi}{a^2}C_{\pi_\phi}-4a^4V'(\phi)C_\phi-2\pi_\mathbf{k} C_{\pi_{\bf k}}-2\omega^2 v_\mathbf{k} C_{v_{\bf k}},\\
\{C_{\pi_\phi},C_{\pi_{\bf k}}\} &=& 2\omega^2 v_\mathbf{k} C_{\pi_\phi}-2a^4V'(\phi)C_{\pi_{\bf k}},\\
\{C_{\pi_\phi},C_{v_{\bf k}}\} &=& 2\pi_\mathbf{k} C_{\pi_\phi}-2a^4V'(\phi)C_{ v_{\bf k}},\\
\{C_\phi,C_{\pi_v}\}&=&2\omega^2 v_\mathbf{k} C_\phi + \frac{2\pi_\phi}{a^2}C_{\pi_v},\\
\{C_\phi,C_{v_{\bf k}}\}&=&-2\pi_\mathbf{k} C_\phi + \frac{2\pi_\phi}{a^2}C_{v_{\bf k}},\\
\{C_{\pi_{\bf k}},C_{v_{\bf k}}\} &=& 2 G \pi_a C_{\pi_a} + \frac{2}{a^3}\left(\pi_\phi^2-4a^6V(\phi)\right)C_a + \frac{2\pi_\phi}{a^2}C_{\pi_\phi}-2a^4V'(\phi)C_\phi-4\pi_\mathbf{k} C_{\pi_{\bf k}}-4\omega^2 v_\mathbf{k} C_{v_{\bf k}}.
\end{eqnarray*}
}

\section{The Dirac brackets}

In this appendix we verify that the matrix that must be defined to construct
the Dirac brackets when imposing the gauge conditions \eqref{gaugefixing1}--\eqref{gaugefixinglast}
on the first-class constraint system \eqref{constraintc3}--\eqref{constraintc7}
is invertible in the general case. Let us define the whole set of constraints by $\Phi_I=\phi_I$,
for $I=1,2,3,4,5$, and $\Phi_I=C_{I-3}$, for $I=6,7,8,9,10$.
Then the Poisson bracket matrix is defined by $M_{IJ}:=\{\Phi_I,\Phi_J\}$, and its determinant is given by,
\begin{eqnarray*}
 det(M)&=& -\frac{1}{64} \left\{\hbar^5+4 \hbar^3
   \left[(\Delta(\pi_\textbf{k} v_\textbf{k}))^2+(\Delta(\pi_\phi \phi))^2-(\Delta \pi_\textbf{k})^2 (\Delta v_\textbf{k})^2+2 \Delta(\pi_\phi v_\textbf{k})
   \Delta(\phi \pi_\textbf{k})
\right.\right.   \\ &-&\left.\left.2 \Delta(\phi v_\textbf{k})
   \Delta(\pi_\phi \pi_\textbf{k})
- (\Delta \phi)^2(\Delta \pi_\phi)^2\right]
+16 \hbar
   \left[(\Delta(\pi_\phi \phi))^2
   (\Delta(\pi_\textbf{k} v_\textbf{k}))^2
   \right.\right.\nonumber\\ &-& \left.\left.   
   (\Delta \phi)^2 (\Delta \pi_\phi)^2
   (\Delta(\pi_\textbf{k} v_\textbf{k}))^2
   +2 (\Delta \pi_\phi)^2 \Delta(\phi \pi_\textbf{k}) \Delta(\phi v_\textbf{k}) \Delta(\pi_\textbf{k} v_\textbf{k})
      \right.\right.\nonumber\\ &-& \left.\left.
   2
   \Delta(\phi v_\textbf{k}) \Delta(\pi_\phi \pi_\textbf{k}) \Delta(\pi_\phi \phi)
   \Delta(\pi_\textbf{k} v_\textbf{k})
   +(\Delta(\phi v_\textbf{k}))^2
   (\Delta(\pi_\phi \pi_\textbf{k}))^2
      \right.\right.\nonumber\\ &-& \left.\left.
   (\Delta \pi_\textbf{k})^2 (\Delta \pi_\phi)^2
   (\Delta(\phi v_\textbf{k}))^2+(\Delta(\pi_\phi v_\textbf{k}))^2
   \left((\Delta(\phi \pi_\textbf{k}))^2
   -(\Delta \pi_\textbf{k})^2
   (\Delta \phi)^2\right)
         \right.\right.\nonumber\\ &-& \left.\left.   
   (\Delta v_\textbf{k})^2
   \left((\Delta \phi)^2 (\Delta(\pi_\phi \pi_\textbf{k}))^2-2
   \Delta(\phi \pi_\textbf{k}) \Delta(\pi_\phi \phi)
   \Delta(\pi_\phi \pi_\textbf{k})
   +(\Delta \pi_\textbf{k})^2(\Delta(\pi_\phi \phi))^2
               \right.\right.\right.\nonumber\\ &+&\left.\left.\left.
   (\Delta \pi_\phi)^2
   \left((\Delta(\phi \pi_\textbf{k}))^2-(\Delta \pi_\textbf{k})^2
   (\Delta \phi)^2\right)\right)
   -2 \Delta(\pi_\phi v_\textbf{k})
   ((\Delta(\pi_\textbf{k} v_\textbf{k}) \Delta(\phi \pi_\textbf{k})
   \right.\right.\nonumber\\ &-&\left.\left.
   (\Delta v_\textbf{k})^2
   \Delta(\phi v_\textbf{k})) \Delta(\pi_\phi \phi)+\Delta(\pi_\phi \pi_\textbf{k})
   (\Delta(\phi \pi_\textbf{k}) \Delta(\phi v_\textbf{k})-\Delta(\pi_\textbf{k} v_\textbf{k})
   (\Delta \phi)^2))\right]\right\}^2.
\end{eqnarray*}
This is a complicate polynomial expression, which will be nonvanishing for generic
values of the different moments. In this general case, the Dirac brackets can be
properly defined and the chosen gauge is thus valid.

Nonetheless, this expression might be vanishing for certain special combination of moments.
It is not possible to find all the roots of the determinant, but there is one particular
and interesting case, for which this determinant vanishes. This happens for the case of a product
state that saturates the Heisenberg uncertainty principle of every degree of freedom.
In terms of the moments, this implies that the different degrees of freedom are independent,
and thus the correlations $\Delta(\phi \pi_\textbf{k})$, $\Delta(\phi v_\textbf{k})$,
$\Delta(\pi_\phi \pi_\textbf{k})$, $\Delta(\pi_\phi v_\textbf{k})$ are zero; and, in addition,
the fluctuations and correlations of each degree of freedom must be related by,
\begin{equation*}
 (\Delta \phi)^2(\Delta \pi_\phi)^2- \Delta (\phi \pi_\phi)^2=\frac{\hbar^2}{4},\qquad
 (\Delta \phi_{\bf k})^2(\Delta v_{\bf k})^2- \Delta(\phi_{\bf k} v_{\bf k})^2=\frac{\hbar^2}{4}.
\end{equation*}
Nevertheless, in this case the gauge above is also valid since, it can be shown that
the null space of the matrix $M$ acts trivially on the physical variables and thus
does not generate any Poisson flow. The degeneracy of the matrix in this special case
of saturation of the uncertainty principle seems to be a generic
feature of this formalism, since it also happens for other models studied in the
literature, see for instance \cite{BT09}, \cite{BHT11}. In fact, in \cite{BT09}
there is a typo and the determinant computed in Section 3.2. is also vanishing for
saturation of the uncertainty relation, but this does not affect the general conclusion
since the same argument as here applies for the validity of the imposed gauge.


\section*{Acknowledgments}
The authors would like to thank M. Bojowald and A. Tsobanjan for correspondence,
and L. Chataignier, C. Kiefer, and M. Kr\"amer for comments on the manuscript.
This work has been supported by Project FIS2017-85076-P (MINECO/AEI/FEDER, UE)
and Basque Government Grant No.~IT956-16.



\end{document}